\documentclass[useAMS,usenatbib]{mn2e}
\usepackage{graphicx}
\usepackage{amsmath}
\usepackage[usenames]{color}

\usepackage{pifont}
\usepackage{amssymb}

\def\deltaL{$\Delta\ln\mathcal{L}$}

    \title[Searching for stellar pulsations in M-dwarfs]
    {High-cadence spectroscopy of M-dwarfs~-~II.\\ Searching for stellar pulsations with HARPS}
   
     \author[Berdi\~nas et. al.]{Z.M. Berdi\~nas$^{1}$,\thanks{E-mail: zaira@iaa.es}
	C. Rodr\'iguez-L\'opez$^{1}$,
	P.J. Amado$^{1}$,	
   	G. Anglada-Escud\'e$^{2}$,
	J.R. Barnes$^{3}$,
     \newauthor{J. MacDonald$^{4}$,
	M. Zechmeister$^{5}$, and L. F. Sarmiento$^{5}$
	}\\
	$^{1}$Instituto de Astrof\'isica de Andaluc\'ia -CSIC, Glorieta de la Astonom\'ia S/N, E-18008 Granada, Spain\\
 	$^{2}$School of Physics and Astronomy, Queen Mary University of London, 327 Mile End Rd., London, E1 4NS, UK\\
	$^{3}$Department of Physical Sciences, The Open University, Walton Hall, Milton Keynes, MK7 6AA, UK\\
	$^{4}$Department of Physics and Astronomy, University of Delaware, Newark, DE 19716, USA\\
	$^{5}$Institut f\"{u}r Astrophysik, Georg-August-Universit\"{a}t G\"{o}ttingen, Friedrich-Hund-Platz 1, 37077 G\"{o}ttingen, Germany
	}
\begin{document}
    \date{Accepted YYYY MM DD. Received MM DD, YYYY; in original form MM DD, YYYY}
\pagerange{\pageref{firstpage}--\pageref{lastpage}} \pubyear{2015}
\maketitle
\label{firstpage}

\begin{abstract}

Stellar oscillations appear all across the Hertzsprung-Russell diagram. Recent theoretical studies support their existence also in the atmospheres of M dwarfs. These studies predict for them short periodicities ranging from 20~min to 3~h. Our Cool Tiny Beats (CTB) programme aims at finding these oscillations for the very first time. With this goal, CTB explores the short time domain of M dwarfs using radial velocity data from the HARPS-ESO and HARPS-N high-precision spectrographs. Here we present the results for the two most long-term stable targets observed to date with CTB, GJ~588 and GJ~699 (i.e. Barnard's star). In the first part of this work we detail the correction of several instrumental effects. These corrections are specially relevant when searching for sub-night signals. Results show no significant signals in the range where M dwarfs pulsations were predicted. However, we estimate that stellar pulsations with amplitudes larger than $\sim0.5\,\mathrm{m\,s}^{-1}$ can be detected with a 90\% completeness with our observations. This result, along with the excess of power regions detected in the periodograms, open the possibility of non-resolved very low amplitude pulsation signals. Next generation more precise instrumentation would be required to detect such oscillations. However, the possibility of detecting pulsating M-dwarf stars with larger amplitudes is feasible due to the short size of the analysed sample. This motivates the need for completeness of the CTB survey.
\end{abstract}


\section{Introduction}
Our Galaxy is mostly populated by low mass stars. In particular, more than 72\% of our stellar neighbours are main sequence stars with masses ranging from 0.08 up to $0.60\,\mathrm{M}_{\odot}$ \citep{henry2006}. A better modelling of the fundamental physical properties of these abundant low mass stars is important, not just to better understand the stars themselves, but also to address fundamental questions such as their contribution to the total mass of our Galaxy; a parameter with cosmological implications that still requires a precise derivation of the mass$-$luminosity relation for different metallicities.

The theory of asteroseismology, which studies the seismology of the stars, has been demonstrated to be a powerful tool to refine the models that describe the stellar structure and evolution (i.e. the combination of seismic data with classical astronomical observations makes possible to calibrate the theoretical models). Additionally, using the asteroseismology we can infer the properties of the stellar interiors. This allow us to calculate at an unprecedented level of accuracy not only the main physical parameters of the star, but also of its hosted planets. Consequently, parameters such as the planet mass, radius, or even the average surface temperature or the planet orbit obliquity can be measured at a very high precision level \citep[e.g. Kepler~-~10, Kepler 56 or Kepler~-~419;][]{fogtmannschulz2014, huber2013, dawson2014}. 

Recently, a new theoretical study from \citealt{rodriguezlopez2014} (from now on RL14) predicts that low-mass M-dwarf stars (0.3$-$0.6$\mathrm{M}_{\odot}$) have the potential to pulsate. Consequently, M dwarfs may have mechanisms able to start, drive and maintain stellar pulsations in their interior. This opens the possibility for the application of asteroseismic tools to M dwarfs, but first the pulsations have to be observationally confirmed. 

RL14 predicts two main regions where M dwarfs may be able of maintaining pulsations. While one comprises young pre-main sequence stars, the other is formed by M dwarfs on the main sequence. For the latter, which corresponds to the M dwarfs observed with our observational programme, RL14 predict pulsation periods in the 20~min up to 3~h range (8 to 72~d$^{-1}$ in frequency). Two driving mechanisms are at work to maintain the oscillations: i) the $\epsilon$ mechanism, caused by He$^{3}$ burning, that works on its own in the 20$-$30~min range for completely convective models (0.20$-$0.30$\mathrm{M}_{\odot}$); and ii) the so-called ``flux-blocking mechanism'', that acts periodically blocking the radiative flux at the tachocline (i.e. the transition layer from the radiative interior and the convective exterior), and that is the main driver of the pulsations in the whole 20~min to 3~h range for models with masses in the 0.35$-$0.60$\mathrm{M}_{\odot}$ range. RL14 is an extension of \citealt{rodriguezlopez2012} to include the excitation of not just the fundamental radial mode, but also non-radial and non-fundamental p-modes and g-modes. Such a range of predicted periods gives us a starting point to start the search for
stellar pulsations in main sequence stars. 

Even though the theory works well in predicting the expected periods, the current existing linear oscillation codes cannot predict the amplitudes of the oscillations. So far, photometric campaigns have only been able to establish upper limits on the amplitudes. Indeed, \cite{rodriguez2016} have recently performed an extensive and exhaustive analysis of 87 M dwarfs observed at high-precision and short-cadence (1~min) with the \emph{Kepler} spacecraft. Although they did not find any significant signal in the 10-100~$\mathrm{d}^{-1}$ range, they set up a new photometric detection threshold of tens of $\mu\rm mag$. However, this low photometric limit does not imply that stellar pulsations are undetectable using high-precision radial velocity (RV) spectrographs. In fact, radial and non-radial pulsation modes detected in both, photometric and spectroscopic observations for other spectral types, indicate that a signal of $10~\mu\rm mag$ can have a counterpart of $1~\mathrm{m\,s}^{-1}$ in RVs \cite[e.g. $\delta$~Scuti and $\gamma$~Dor oscillators such as FG~Vir, RZ~Cas, 9-Aur, HR~8799, gam~Dor, and HD~49434;][]{zima2006, lehmann2004, zerbi1997, zerbi1999, krisciunas1995, uytterhoeven2008}; an amplitude that is detectable with HARPS-ESO, which has a reported RV precision in the short-term of $0.5\,\rm m\,s^{-1}$ \citep{lovis2010}.

Motivated by this favourable relation we started in 2013 the Cool Tiny Beats project (CTB), which uses the high-precision RV spectrographs HARPS-ESO (hereafter HARPS) and HARPS-N with the goal of detecting stellar pulsations in M dwarfs for the first time. We present the first results of the CTB programme on the search for stellar pulsations. In particular, we present the analysis on the most long-term-stable stars of our sample observed up to date with CTB; GJ~588 and GJ~699 (also known as
Barnard's star). This manuscript is organised as follows: Section~\ref{sec:observations} introduces the CTB survey,
observations and data reduction. In Section~\ref{sec:systematics} we discuss and correct the data from instrumental effects.
Section~\ref{sec:analysis} is the main part of this study comprising the search for periodic signals embedded in the GJ~588 and
GJ~699 time-series (subsection~\ref{sec:periodicsignals}), a short discussion about the pulsation modes we could expect on
GJ~588 (subsection~\ref{sec:putativesignal}), and the empirical calculation of HARPS precision limit in the high-cadence domain
(subsection~\ref{sec:detlim}). Finally, in Section~\ref{sec:discussion} we give the main conclusions of this work. 

\section{The CTB programme: Observations and Data Reduction}\label{sec:observations}

	\begin{figure*}
	\centering
	\includegraphics[width=10cm]{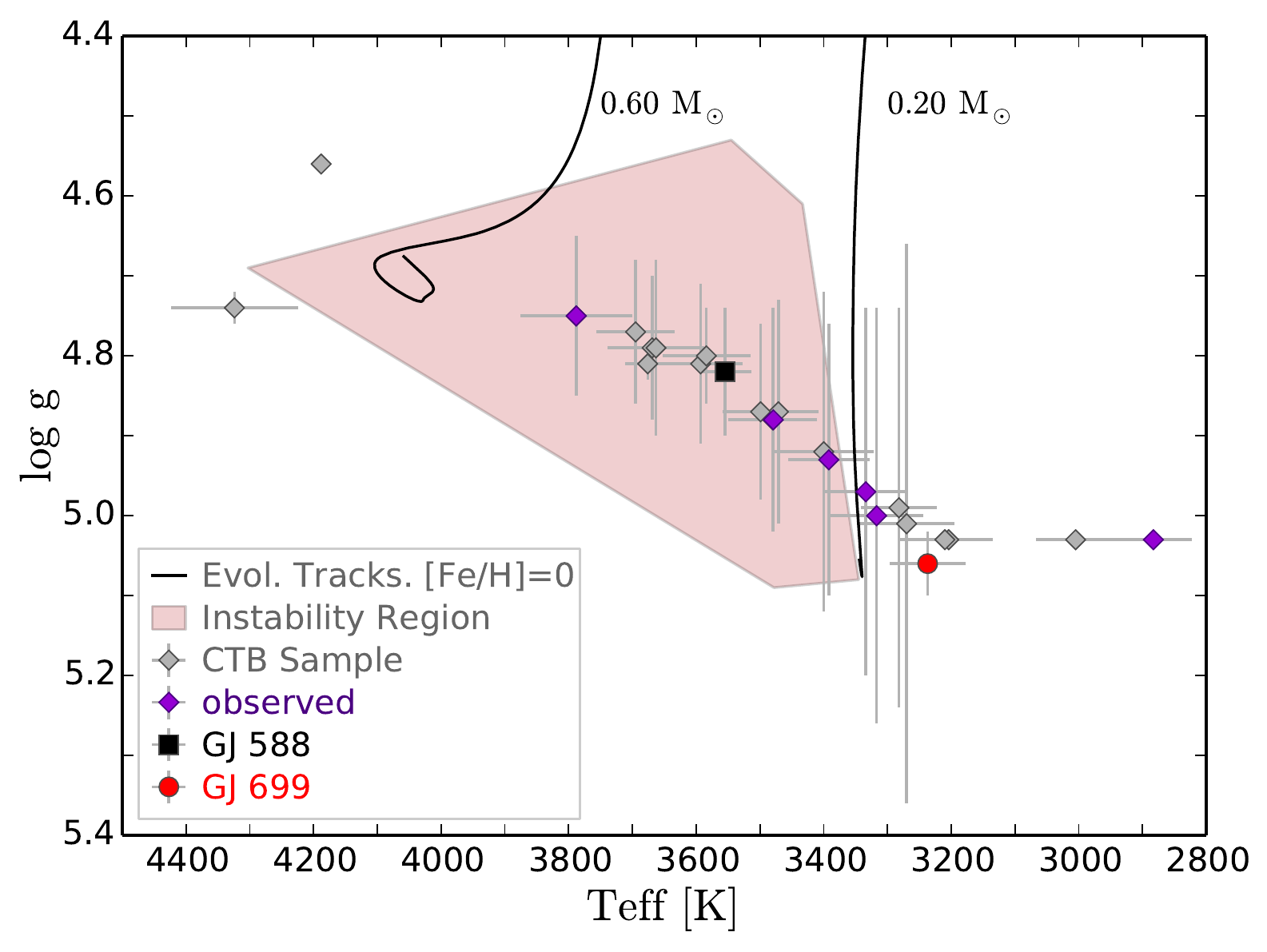}
	\caption{Instability region (pink area) of main-sequence M dwarfs predicted by RL14. The black square and red dot correspond to GJ~588 and GJ~699, respectively. The diamonds indicate other CTB targets, where in purple we highlight those which have been already observed. The black solid lines are 0.20 and 0.60~M$_\odot$ evolutionary tracks (solar metallicity and mixing length parameter $\alpha$=1) delimiting the instability region. Physical parameters come from: \citet{gaidos2014}, \citet{boyajian2012}, \citet{santos2013}, \citet{dressing2013}, and \citet{steffen2013}.}
	\label{fig:ctbsample}
	\end{figure*}

In addition to the high-precision spectroscopy, a continuous time monitoring of the target is essential in asteroseismic
campaigns. With CTB we have been exploring the short time domain of a sample of bright M dwarfs with the scientific
objective of confirming the predicted, but still undetected, stellar pulsations of M dwarfs. In particular, we performed
high-cadence observations, meaning in this case that we continuously monitored the same target with exposures shorter than 20~min during
several consecutive nights. This observational strategy assists also in the robust detection of small planets in warm/potentially habitable orbits around the stars, other of the CTB science goals \citep[e.g.][]{anglada2016a}. In
addition to that, understanding the Doppler variability of the targets is very valuable (even necessary) in the interpretation of the noise sources that can potentially inject false positives in periods of a few days to weeks \citep{anglada2016b}. Compared to the classical planet searches, the unusual high-cadence sampled by CTB allowed us to explore the behaviour of the instrument within the night, identify sources of systematic noise and correct for them in some cases \citep[e.g.~][]{berdinas2016}.

The CTB initial sample was comprised by 25~M dwarfs. The targets were mostly selected to lie within the boundaries of one of the
instability regions defined in RL14, in particular, the one mainly comprised by main sequence M stars. Other criteria were to have demonstrate long-term
Doppler stability ($\Delta_v$(rms)~$<~2.5\,\rm m\,s^{-1}$), low activity levels and slow rotation. Such sample would satisfy the requirements for both the asteroseismology and planetary science cases of CTB. In Figure~\ref{fig:ctbsample} we show the initial CTB sample on a $T_{\rm eff}-\log{g}$ diagram. We have highlighted GJ~588 and GJ~699 (Banard's star), focus of this study, with a black square and a red dot, respectively. For a future CTB Phase II programme we plan to update our sample including more targets on both instability regions predicted by RL14.

	\begin{table}
	\begin{center}
	\caption{GJ~588 and GJ699 stellar parameters.}     
	\label{tab:params}
	\begin{tabular}{lrrl}\\\hline
	Parameters & GJ~588&GJ~699& Refs.\\\hline
	$\rm SpT$   				    	& M2.5		    	& M4		    		& RE95  \\
	$V_{\rm mag}$ 	              	 		& $9.31$           		& $9.51$		   	& KO10 \\
	$\rm Dist.\;[pc]$ 					& $5.93\pm0.05$		& $1.82\pm0.01$	   	& KO10\\
	$P_{\mathrm{rot}}\;[\mathrm{d}]\,^*$		& $61.3\pm6.5$  		& $148.6\pm0.1$		& SU15 \\
	$v\sin{i}\;[\rm km\,s^{-1}]$ 			& $<3.0$			& $<2.5$		   	& RE12 \\
	$\rm [Fe/H]\;[dex]$       			& $0.06\pm0.08$		& $-0.51\pm0.09$		& NE14\\
	$T_{\rm eff}\;[\rm K]$ 				& $3555\pm41$	& $3237\pm60$	    	& GA14\\
	$M\;[\rm M_{\odot}]$ 				& $0.43\pm0.05$		& $0.15\pm0.02$	    	& GA14, BO12\\
	$R\;[\rm R_{\odot}]$ 				& $0.42\pm0.03$		& $0.187\pm0.001$		& GA14, BO12 \\
	$\log{g}\;[\rm cm\,s^{-2}]\,^\dagger$ 		& $4.82\pm0.08$		& $5.040\pm0.005$		&\\\hline
	\end{tabular}
	\end{center}
	References: [BO12]~\cite{boyajian2012}, [GA14]~\cite{gaidos2014}, [KO10]~\cite{koen2010}, [NE14]~\cite{neves2014}, [RE95]~\cite{reid1995}, [RE12]~\cite{reiners2012},  [SU15]~\cite{suarezmascareno2015}.\\ $[*]$~SU15 used also CTB data to calculate the $P_{\rm rot}$ values. $[\dagger]$~The $\log{g}$ values were obtained from this table parameters.
	\end{table}

Barnard's star is an extensively studied M dwarf which does not have any reported
planets and it is known to be stable in the long-term (e.g. \citealt{zechmeister2009b} derived a RV stability of 2.70~$\mathrm{m\,s}^{-1}$ after subtracting the secular acceleration\footnote{The secular acceleration effect is caused by the high proper motion of the nearby stars and results in a linear trend in the RVs.} using UVES data, later \citealt{anglada2012b} reduced the limit
down to 1.23~$\mathrm{m\,s}^{-1}$ using HARPS, and more recently, \citealt{choi2013} found no significant periodic Doppler
signals with amplitudes above $\sim2\,\mathrm{m\,s}^{-1}$ using 25 years of data from the Lick and Keck Observatories). In fact
CTB uses GJ~699 as its RV standard. That is, in addition to the high-cadence observations presented here, we usually took at least one spectrum
of GJ~699 per night when it was observable. For this reason, even when GJ~699 lies in the outer edge of the theoretical
instability region, we decided to include it in this study --the instability region gives us a starting point in the search
for pulsations, but we should not necessarily exclude targets close to its edge since its boundaries were not yet observationally constrained. On the other hand, GJ~588 is both, stable and lies well within the predicted
instability region. We show the relevant stellar parameters of GJ~588 and GJ~699 in  Table~\ref{tab:params}.

The asteroseismology case of CTB requires of high-precision spectroscopy. This is the reason why the programme makes use of HARPS and HARPS-N, the most stable current
instrumentation of this kind. In particular, the data presented here were obtained using the HARPS \'echelle
spectrograph at the 3.6-m telescope at La Silla Observatory. HARPS is a stabilised high-resolution fibre-fed spectrograph which operates
in the visible range (380-680~nm) with a resolving power of R~$\sim110,000$. The reported $0.8-0.9\,\rm m\,s^{-1}$ RV long-term
stability limit \citep{pepe2011}, also on M-dwarfs \cite{anglada2012b}, makes it currently the most suitable spectrograph for
this project. 

\begin{figure}
	\centering
	\includegraphics[width=1\columnwidth]{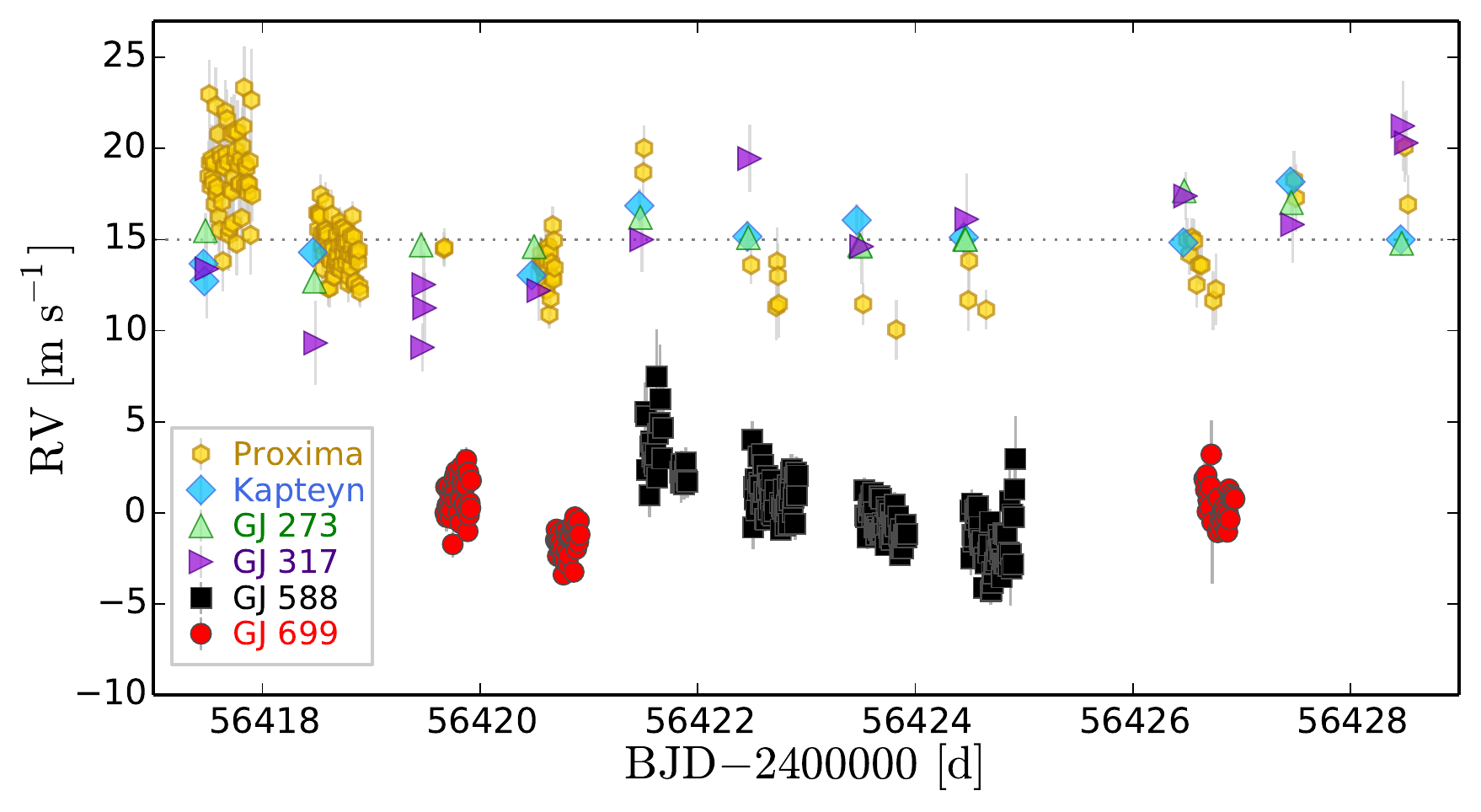}
	\caption{RVs of the CTB May 2013 run from HARPS. Six targets were observed during 11 nights. GJ~588 (black squares) and GJ~699 (red dots) are the focus of this study. The RVs of the other targets (Proxima Centauri -yellow hexagons-, Kapteyn's star -blue diamonds-, GJ~273 or Luyten's star -green up triangles-, and GJ~317 -purple right triangles-) were shifted +15~$\mathrm{m\,s}^{-1}$ for visualization reasons.} 
	\label{fig:may2013rvs}
	\end{figure}

The GJ~588 and GJ~699 high-cadence data were obtained during a CTB observational campaign carried out in May 2013 (see the time series of the targets observed during this run in Figure~\ref{fig:may2013rvs}, the black squares and
red dots account for GJ~588 and GJ~699, respectively). We monitored GJ~588 during four consecutive nights. On the contrary, we observed GJ~699 during three nights taken immediately before and after the GJ~588 observations (two consecutive nights were taken before and one was observed after). In total, we obtained 189 and 108 spectra for GJ~588 and GJ~699, respectively. The exposures times were $\sim600\,\rm s$ for GJ~588 and either 500 or $600\,\rm s$ for GJ~699. Among other observational parameters, we got, respectively for GJ~588 and GJ~699, maximum airmasses of 2.2 and 2.0, mean signal-to-noise ratios (SNR) of 63.1 and 62.8, and mean seeing values of 1.03 and 0.64.

The spectra were extracted and wavelength calibrated with the standard HARPS Data Reduction Software (DRS).  We used the
Template Enhanced Radial velocity Reanalysis software (TERRA) to calculate the RVs, since it has demonstrated to
be more accurate on M dwarfs \citep{anglada2012b}. Additionally, we also calculated a proxy of the mean-line profile of the spectra. In particular, instead of using the cross-correlation functions given by the HARPS pipeline, we calculated them usign a least-square deconvolution technique \citep[LSD;][]{donati1997}. See more details in Section~\ref{sec:coloreffect}.

\section{Instrumental effects correction}\label{sec:systematics}

The analysis of the CTB data revealed both intra-night \citep{berdinas2016} and possibly night-to-night instrumental effects that need to be
mitigated when dealing with high-cadence observations. 

\subsection{Wavelength calibration jumps}

One of the effects that we quickly identified in other CTB data were night-to-night RVs jumps of $\sim0.5-2.0\,\rm m\,s^{-1}$. The stability of
the RVs obtained with TERRA or any other reduction software relies on the calibration of the wavelength, which, in HARPS, is given by a Th-Ar hollow cathode lamp (HCL) measurement taken at the beginning of each night. Since the wavelength solution has also an associated
uncertainty, its random errors are bound to produce night-to-night jumps than can easily reach a $\sim1-2\,\rm m\,s^{-1}$ level.
Moreover, compared to the typical G \& K dwarfs for which HARPS was designed, M dwarf stars have most of the flux and Doppler
information in the redder part of their spectra and hence of the detector. As a result, night-to-night RVs become more sensitive to random errors in a
smaller number of the redder diffraction orders. To calculate the night-to-night offsets in the wavelength solution, we firstly selected one GJ~699 spectrum per night. For nights with several GJ~699 spectra we selected the ones observed early after the beginning of the night ($t_{i}$). Then, we obtained the Doppler drift of the wavelength solution between nights as:
\begin{equation}
	\mathrm{drift}  = c\left(\left<\frac{\lambda_{t_{i}}}{\lambda_{t_{0}}}\right> - 1 \right)\:[\rm{m\,s}^{-1}]\,\,,
	\label{eq:wavesol}
	\end{equation}
\noindent
that is, referenced to the first GJ~699 spectrum of the run ($t_{0}$). Here, $\lambda$ accounts for the wavelength of each pixel in the 22 reddest spectral orders and $c$ is the speed of light.

\begin{figure}
	\centering
	\includegraphics[width=\columnwidth]{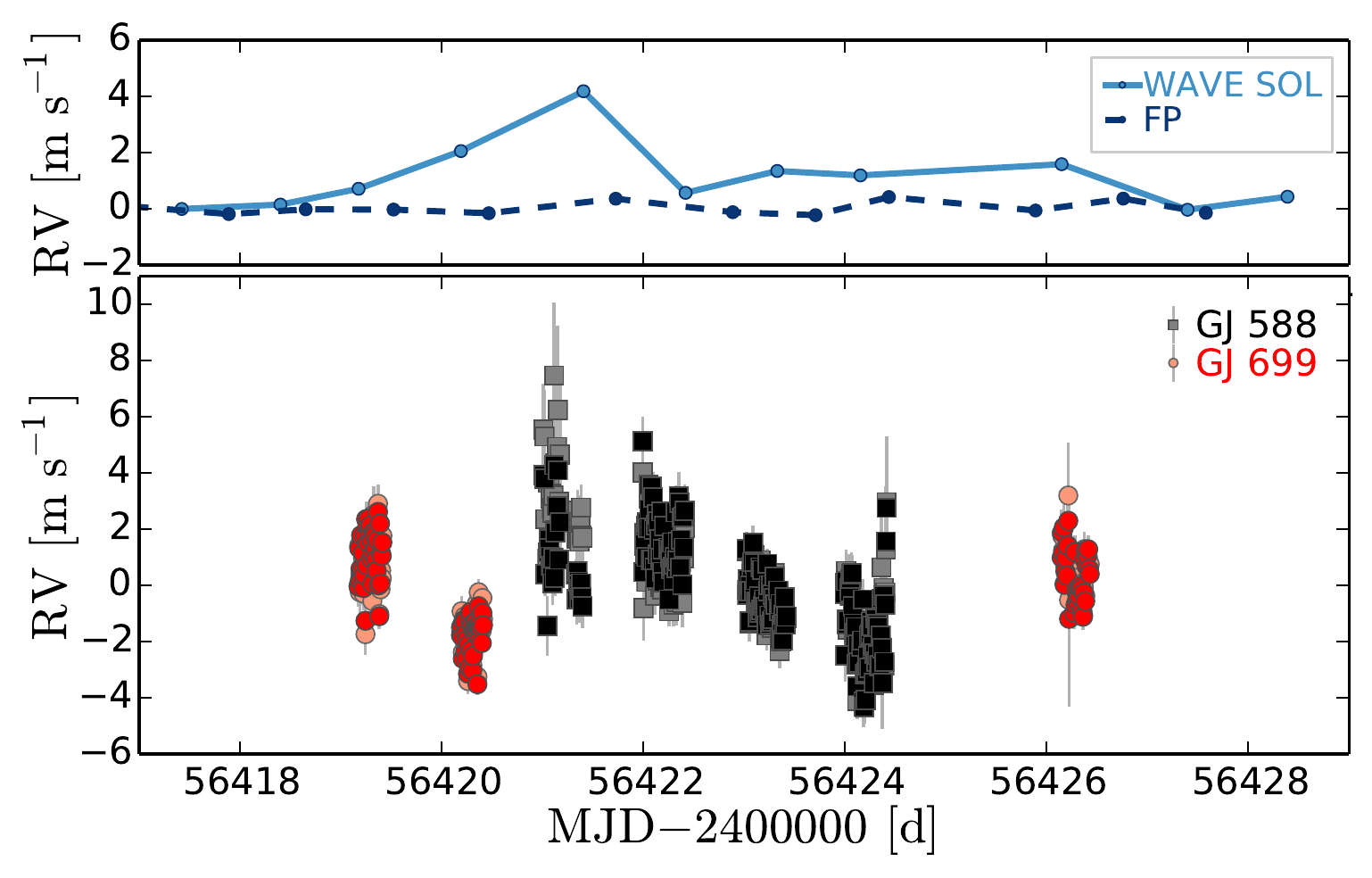}
	\caption{(Upper panel) Comparison of the stability of the wavelength solution (light blue solid line) and Fabry-Perot (dark blue dotted line). The wavelength solution, that relies on the Th-Ar hollow cathode lamp, is less stable. (Bottom panel) The dark color symbols indicate RVs obtained for the same wavelength solution while the light color refer to RVs from individual night calibrations. The squares and dots correspond to GJ~588 and GJ~699, respectively.} 
	\label{fig:wavesol}
	\end{figure}

As a complementary wavelength reference source, a spectrum with a stabilised Fabry-P\'erot (FP) interferometer is also
obtained as part of the calibration procedure executed before the observing night starts\footnote{The FP frames were taken in the fiber B while the spectrum of the Th-Ar hollow cathode lamp was recorded in the fiber A.}. Using TERRA, we computed the RV drift of these FP frames against the FP frame taken in the first epoch. In the upper panel of Figure~\ref{fig:wavesol}, we compare the nightly FP and wavelength
solution drifts. The datapoints of the wavelength solution and FP do not have the same time-stamp (i.e. they do not match in the x-axis of Figure~\ref{fig:wavesol}) because while the FP is recorded before the start of the night, the GJ~699 spectra used to obtain the wavelength solution were observed throughout the night. Nevertheless, we can still compare both drifts because a single calibration procedure is typically used as reference for all the observations of the night. Results indicate that even when the FP measurements show some structure, its corresponding time-series is much more stable ($\Delta_v(\mathrm{rms})=0.22\,\mathrm{m\,s}^{-1}$) than in the case of the wavelength solution ($\Delta_v(\mathrm{rms})=1.16\,\mathrm{m\,s}^{-1}$). Although they are likely to contribute to random noise over long time-scales, the
jumps in the wavelength solution already cause serious issues in the consistency of our time-series in the high-cadence domain (signals
in the P$<2$~d range). For example, we can see in Figure~\ref{fig:wavesol} how a $3.6\,\mathrm{m\,s}^{-1}$ jump in the wavelength
solution causes the drift of the GJ~588 RVs between the first and the second night. As a comparison, the time-series of GJ~588 looks much
flatter if we use, instead of the individual night
calibrations, the wavelength solution of the first night for all the observations (compare black and grey squares in the bottom panel of Figure~\ref{fig:wavesol}). 

This random variability of the
wavelength solution is one of the causes of the spurious $\sim$1-day peaks (and integer fractions of it, like 1/2 and 1/3) that
commonly appear in the periodograms of high-cadence data (e.g. the low frequency excess in the power spectra of fig.~3 and
fig.7 of \citealt{bedding2007} and \citealt{bouchy2005}, respectively). Although we do not expect pulsations at such long
periods, the window function can inject significant correlated noise at other frequencies, which is undesirable. We use the
strategy of using a common wavelength solution for all the run to mitigate this source of noise. The long term stability of the
FP ($>$~week) has not yet been established (Pepe \& Lovis, \emph{priv. comm.}) so we advise against using this technique to improve the
consistency of time-series with a time-span longer than a few days.

\subsection{Charge Transfer Inefficiency}
Doppler shifts have been reported to correlate with the signal-to-noise ratio (SNR) of the observations \citep{bouchy2009}.
This is in part due to the Charge Transfer Inefficiency (CTI) effect. The CTI gets worse at low fluxes (e.i. at low SNRs) and it is associated
with an inefficient transference of charge between adjacent pixels during the readout process in charge-coupled devices (CCD).
The CTI can produce effective changes in the position and the shape of the HARPS spectral lines \citep{locurto2012, zhao2014},
and thus, cause RV offsets of several $\mathrm{m\,s}^{-1}$\citep{bouchy2009} for measurements with SNR below $30-40$. \cite{bouchy2009}
proposed a method to assess the charge lost in the pixels of the SOPHIE spectrograph \citep{perruchot2008}.  Using similar
methods, the raw frames of HARPS-N are corrected for CTI since 2013. However, such a correction was not implemented in HARPS
(Lovis, C., Pepe, F., priv. comm.). 

\begin{figure}
	\centering
	\includegraphics[width=\columnwidth]{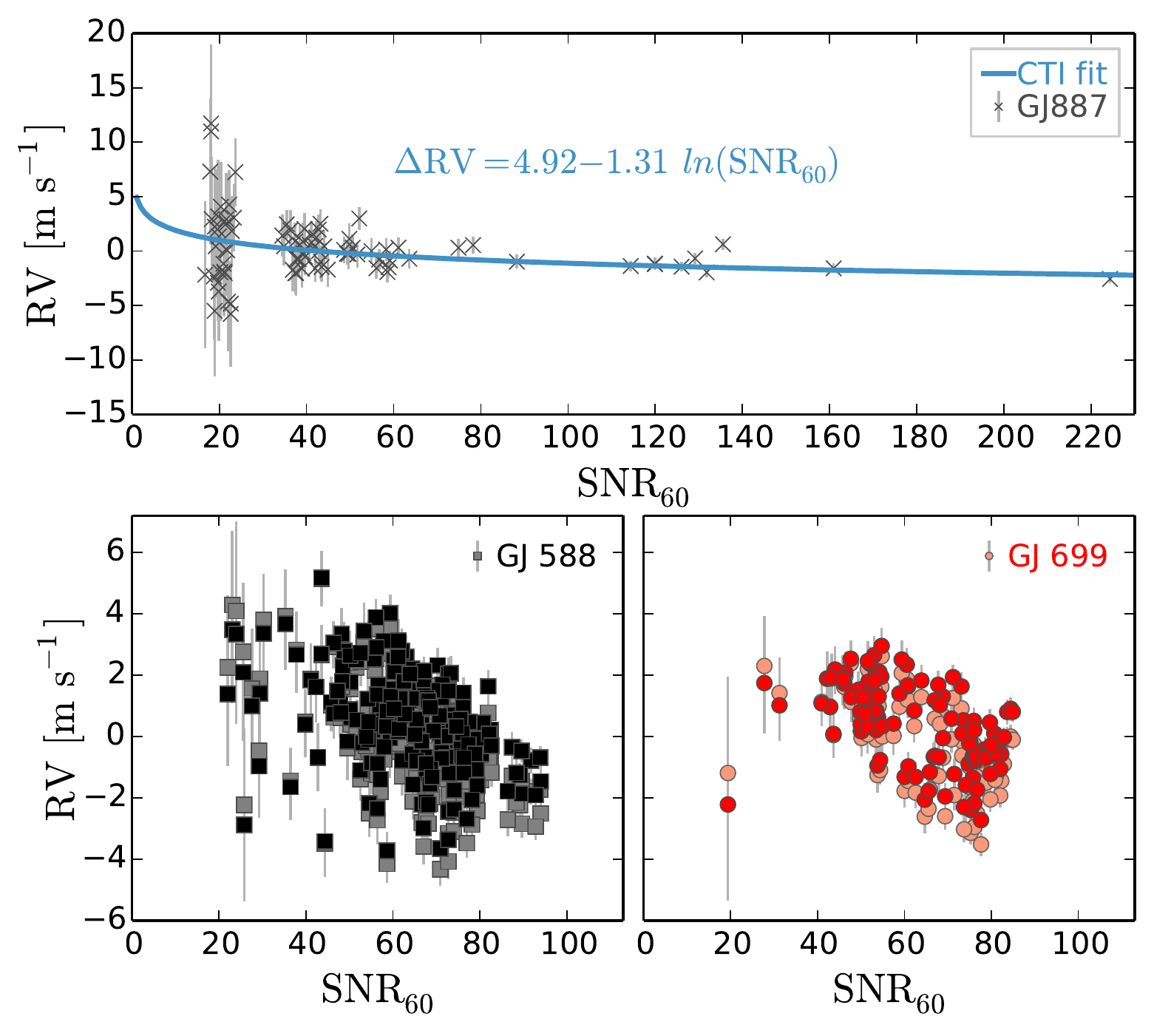}
	\caption{(Upper panel) GJ~887 RVs obtained at different exposure times (from 10 up to 600~s), i.e. for a range of SNRs. This experiment was done to monitor the CTI effect that distorts the RV at low SNR and to get the CTI calibration function (blue solid line and function). (Bottom panels) The light symbols are the original RVs of GJ~588 (left; squares) and GJ~699 (right; dots), while the dark markers correspond to their CTI-corrected counterparts.} 
	\label{fig:cti}
	\end{figure}

To mitigate CTI effects for our campaign, we implemented an empirical
post-processing  correction: during a CTB campaign carried out in December 2014 we observed GJ~887, a very bright M2V spectral type star,
with different exposures times, i.e. at different SNRs. Similarly to what \cite{santerne2012} did for SOPHIE, we fit a relation between RV and SNR which is valid for, at least, M dwarf stars observed with HARPS. This is:
	\begin{equation}
	\Delta \mathrm{RV}= 4.92 -1.31 \ln{\mathrm{SNR}_{\mathrm{60}}}\:[\rm{\,m\,s}^{-1}], 
 	\label{eq:cti}
	\end{equation} 
\noindent
where $\mathrm{SNR}_{\mathrm{60}}$ refers to the SNR measured in the spectral order 60.  In Figure~\ref{fig:cti}, we show the CTI
empirical calibration function in the upper panel and the corrected RVs in the lower part. There are other effects that cause the RVs to correlate with the SNR. An example is the ``the color effect'' outlined in \cite{bourrier2014}. Contrarily to the CTI, which only affects measurements with SNR below $30-40$, these effects can cause trends for the whole range of SNRs as we see in the lower panels of Figure~\ref{fig:cti}.

\subsection{Seeing Effect}
\begin{figure}
	\centering
	\includegraphics[width=\columnwidth]{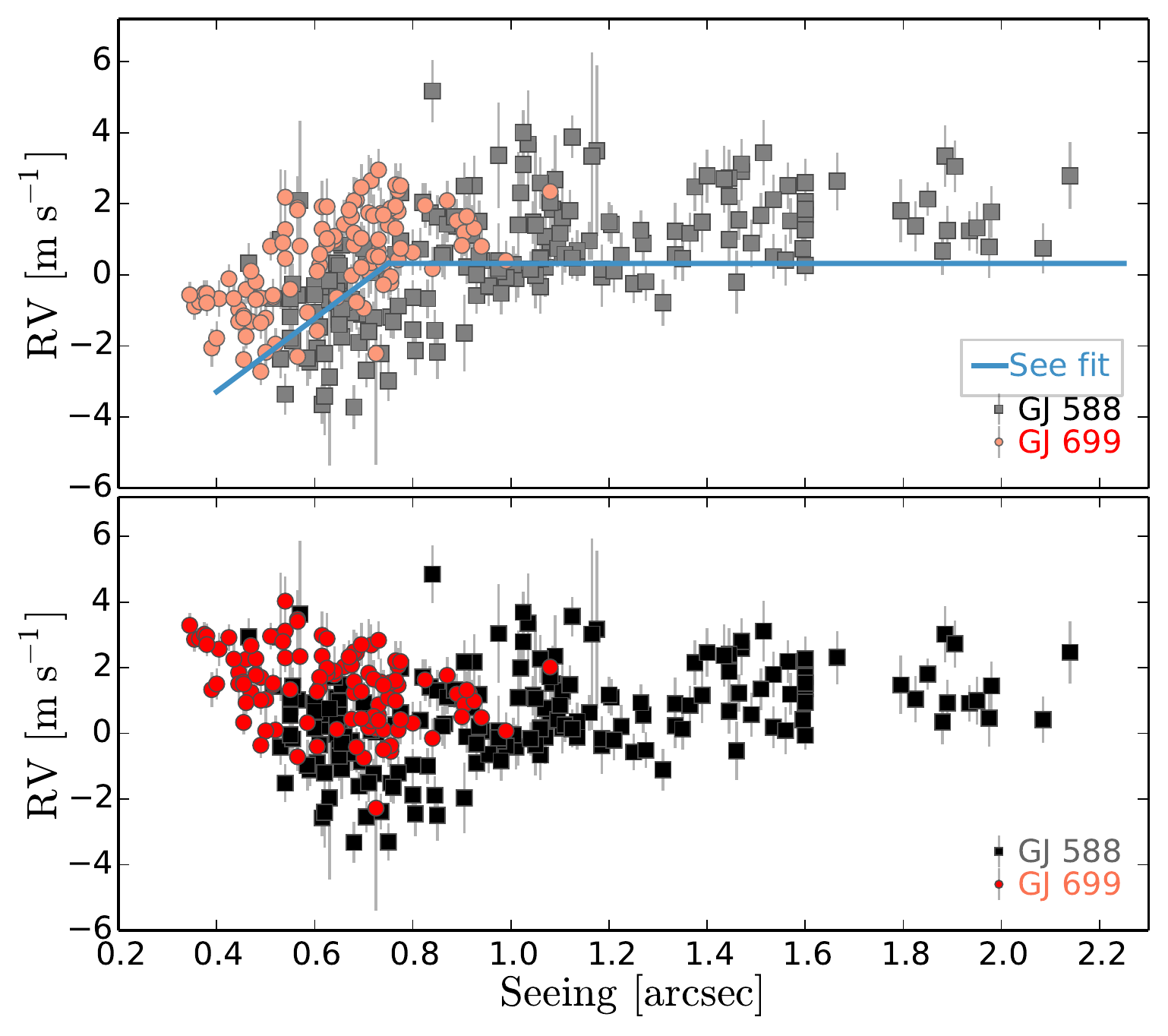}
	\caption{Seeing effect. (Upper panel) GJ~588 (grey squares) and GJ~699 (light red dots) RVs uncorrected from the``seeing effect''. The RVs decrease for the datapoints observed at seeing values lower than the HARPS fibre size, i.e. 1~arcsecond. Based on dispersion criteria, we only applied a linear fit to the RVs below 0.75~arcseconds. Corrected series are shown at the lower panel with dark colours (black for GJ~588 and red for GJ~699).}
	\label{fig:seeingeffect}
	\end{figure}

We observed that when atmospheric conditions were excellent (i.e. seeing $<$ 1 arsecond), the
Doppler measurements of several stars were correlated with lower values of seeing (see Figure~\ref{fig:seeingeffect}). Such effect, known as the ``seeing effect", was well described by \cite{boisse2010a,
boisse2010b}, and it was pointed out as the main limiting factor of the SOPHIE spectrograph \citep{perruchot2008}. The effect can be understood as vignetting of the telescope pupil. \cite{boisse2010a} explained how this vignetting translates into light pattern variabilities at the output of the optical fibre linking the telescope and the spectrograph (in particular, into changes of the  far field image of the fibre); and how this finally produces small shifts of the spectral lines (i.e. of the RVs). HARPS has a double image scrambler to stabilise the image and to homogenise the illumination. Since this system interchanges the fibre near and far fields, vignetting the pupil results equivalent to reduce the size of the image of the star.  That is, equivalent to have good seeing conditions. When the seeing is below 1~arcsecond the telescope image is sharper than the HARPS fibre width, and
thus, if the scrambling is not perfect, the far field changes cause RV shifts up to $\sim3\,\mathrm{m}\,\mathrm{s}^{-1}$ \citep{boisse2010a}.

In 2013, \cite{bouchy2013} showed how to partly correct the ``seeing effect'' of SOPHIE. They demonstrated that using linking fibres with octagonal-shaped cores, instead of circular ones, improves the scrambling efficiency and the resulting RV precision by a factor of $\sim 6$. In May 2015 an octagonal fibre link was introduced in HARPS. However, the data used in this
study were obtained before this instrumental update. Therefore, we applied a post-processing empirical correction based on our
observations. 

We used as a proxy for the seeing the value given in the data headers which is obtained as the FWHM of the acquisition image taken
by the guider camera. Then, we fitted and subtracted a linear function to the observations with seeing $<$~0.75~arcseconds (see
Figure~\ref{fig:seeingeffect}). This value is below the HARPS fibre width (1~arcsecond) and was chosen in terms of the variability observed in the May 2013 data. Moreover, our 0.75-arcsecond cut-off resulted to be in good agreement with the 0.7~arcseconds quality image given by the 3.6-m
telescope where HARPS is installed \citep{boisse2010a}. 

\subsection{The SED normalisation effect}\label{sec:coloreffect}

Our previous studies using CTB high-cadence data revealed systematic effects within the night in HARPS-N \citep{berdinas2016}.
Such systematic effects, dubbed ``SED normalisation effect" (where SED stands for spectral energy distribution), consist in wavelength dependence of the flux losses caused by
small illumination changes at the fibre entrance. 

\begin{figure}
	\centering
	\includegraphics[width=\columnwidth]{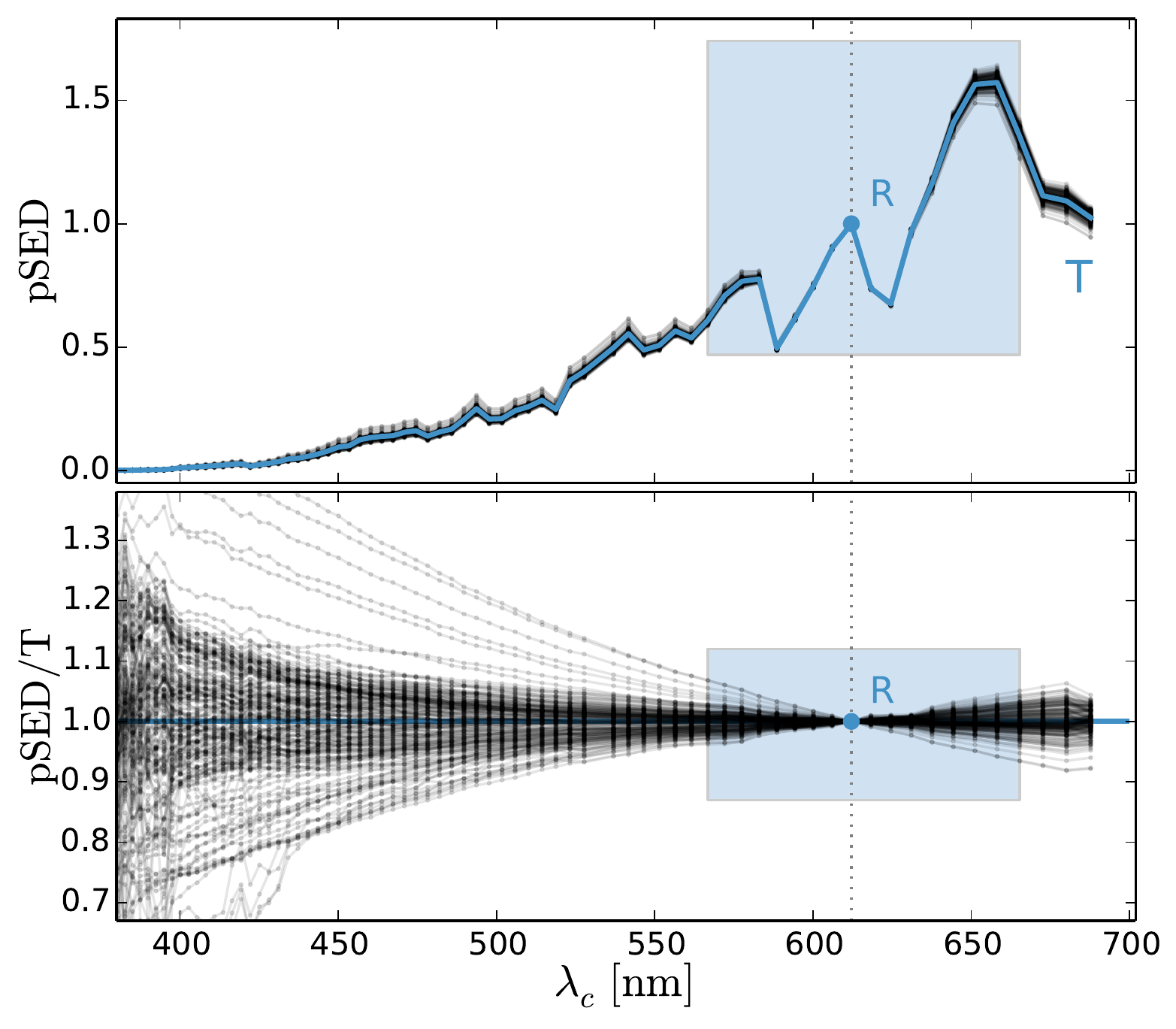}
	\caption{(Upper panel) GJ~588 pseudo Spectral Energy Distribution (or pSED), calculated as the sum of the flux at each spectral order normalised by the flux at order-60. The pSED is plotted versus $\lambda_c$, the central wavelength of each spectral order. R indicates the central wavelength at order-60. (Lower panel) The pSEDs normalised by T, the averaged pSED of the run (see blue line in the upper panel). The index $\kappa$ measures the relative changes in the slope of the pSEDs. It is calculated for the range of wavelengths within the blue area (see \citealt{berdinas2016} for more details).}
	\label{fig:sed}
	\end{figure}

After calculating the pseudo-SED function following \cite{berdinas2016}, we detected wavelength and time dependencies in the flux distributions of GJ~588 and GJ~699 (see Figure~\ref{fig:sed}). Thus, the HARPS-DRS pipeline does not account for the ``SED normalisation effect'', neither for HARPS nor HARPS-N. This implies that the mean-line profile proxies given by the DRS (i.e. the cross-correlation function, or CCFs) are uncorrected, and so the indices derived from them.

\begin{figure}
	\centering
	\includegraphics[width=\columnwidth]{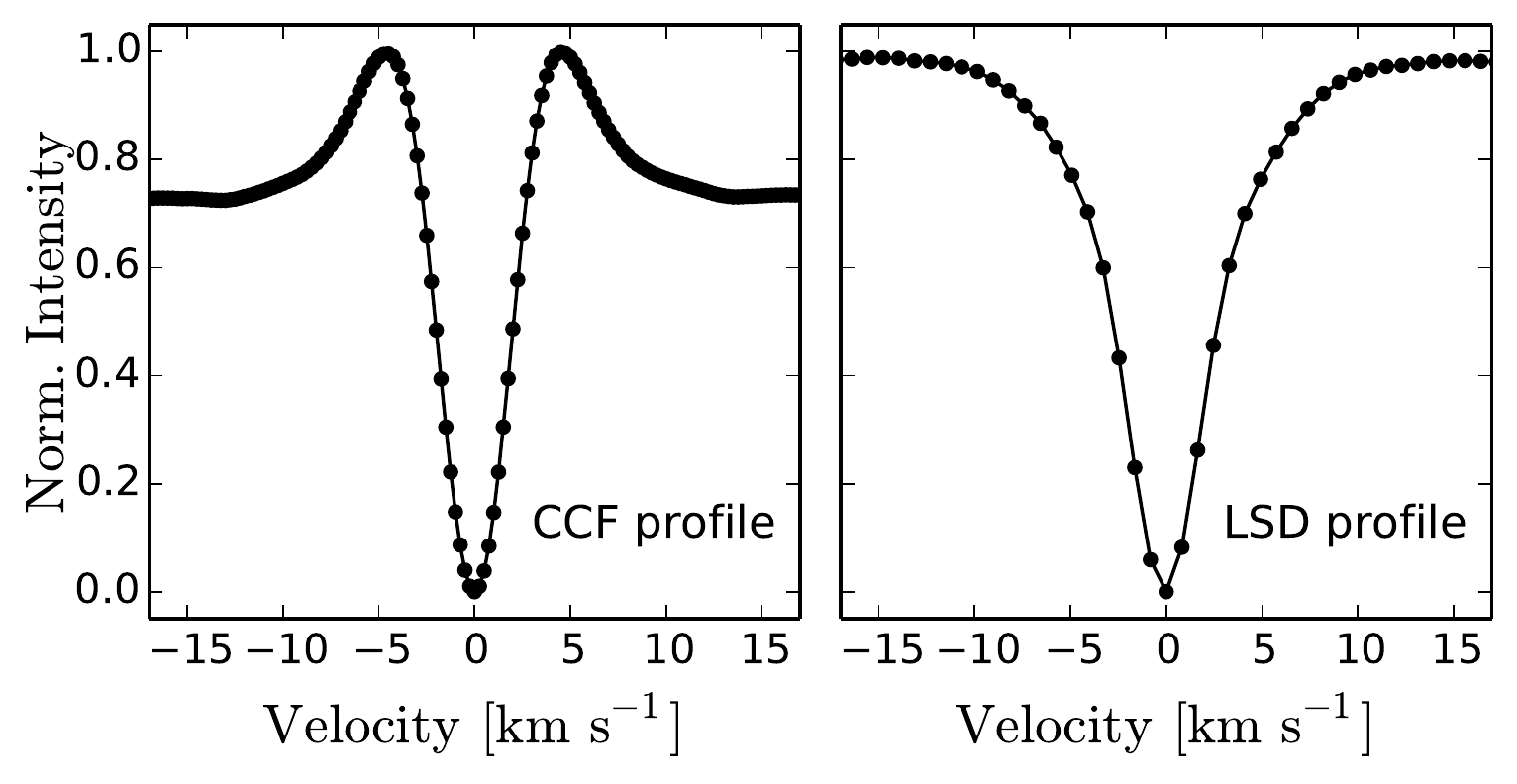}
	\caption{Comparison of a CCF profile obtained from the DRS pipeline (left panel) and a profile obtained with a least-square deconvolution (LSD) approach (right panel). The profiles correspond to the same GJ~588 spectrum, and were shifted to zero-velocity and normalised by its minimum intensity for a better visualization.}
	\label{fig:lsdccf}
	\end{figure}

Thus, in order to measure uncontaminated mean-line profiles we have to correct their variable slope. To this end, we re-scaled the flux at each order and epoch to match that of our template spectrum (with highest SNR). Once we corrected the spectra from the ``SED normalisation effect'', we calculated the mean-line-profile with a LSD approach \citep{donati1997} as outlined in \cite{barnes1998,barnes2012}. Compared to the CCFs from the DRS, our LSD approach resulted in smoother mean-line profiles. As a proof of that, in Figure~\ref{fig:lsdccf} we show a CCF and a LSD mean-line profile for the same GJ~588 spectrum. The side lobes present in the CCF profile\footnote{Side lobes are generally caused by blending between nearby lines. This is typical of cool stars which are crowded with lines that distort the star's continuum.} increase the uncertainty of the Gaussian fit as well as of any other parameters derived from it, such as the full-width-at-half-maximum (FWHM; the line width), the RV (the line centroid), and the bisector (BIS; the line asymmetry).

The LSD method consists in finding the convolution kernel applied to a list of lines to reproduce the observed spectrum in a least-squares sense, thus naturally accounting for the line blends. To create such a list of lines, we used a co-added high-SNR spectrum generated with all the observations. This processing consists in matching the continuum and Doppler shift of each spectrum to the highest SNR spectrum, and then co-adding all of them. When selecting lines for the LSD, chromospheric emission lines (such as H$_\alpha$ and NaI D1 and D2 lines) or any telluric lines deeper than 0.05 compared to the normalised continuum are excluded. The LSD profile is obtained using only the reddest HARPS apertures (from order 32 to 72) and it is sampled at $0.821\,\rm km\,s^{-1}$, which matches the effective average pixel size of HARPS (the DRS oversamples the pixel size using velocity steps of $0.25\,\rm km\,s^{-1}$). 

\begin{figure}
	\centering
	\includegraphics[width=\columnwidth]{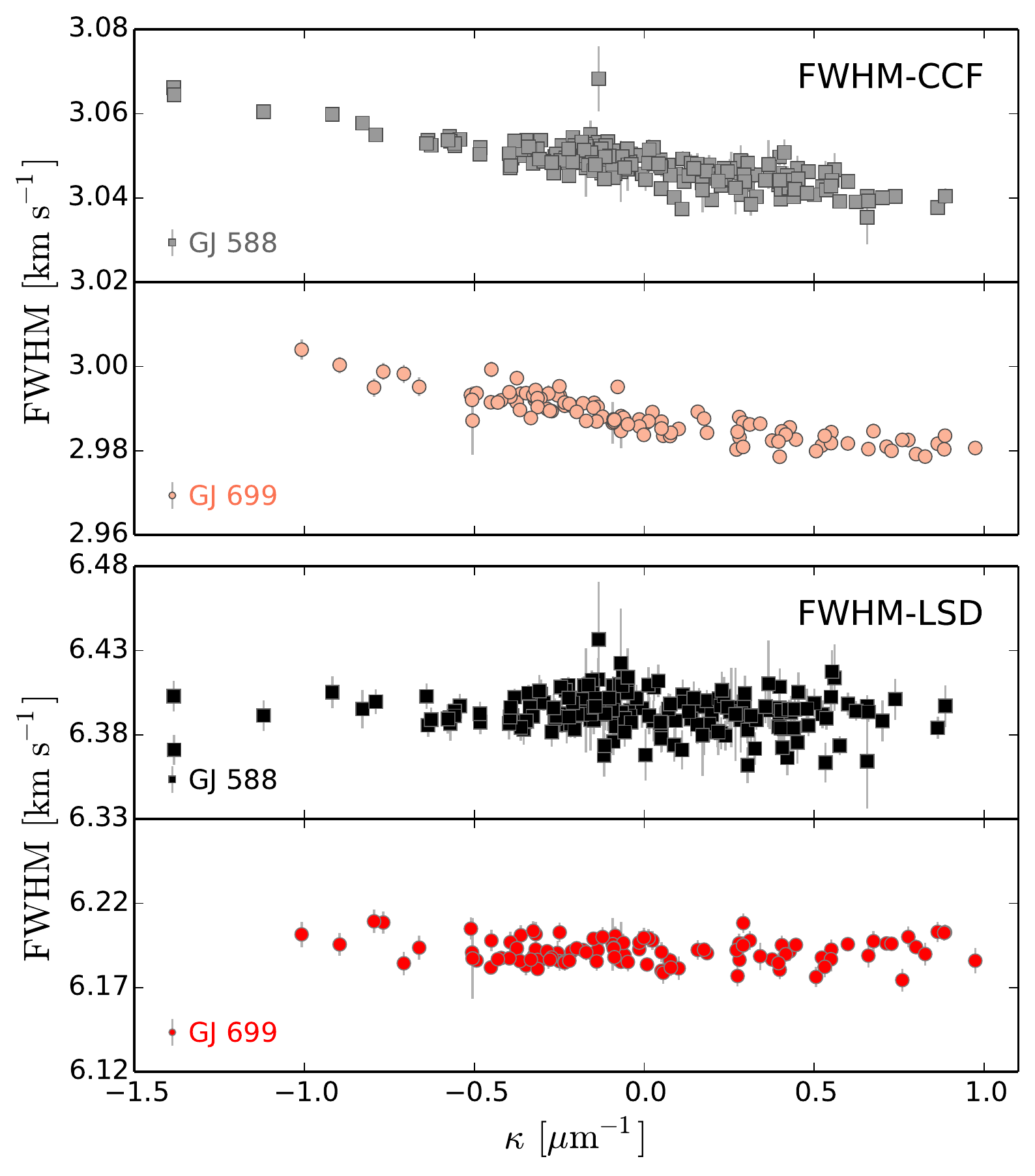}
	\caption{SED normalisation effect. FWHM-CCF (upper panels; light colours) and FWHM-LSD indices (lower panels; dark colours) plotted versus $\kappa$, that accounts for flux variabilities of the SED. The squares and dots correspond to GJ~588 and GJ~699, respectively. The FWHM-CCF and the FWHM-LSD correspond respectively to the full width at half maximum of the Gaussian function fitted to the cross-correlation functions by the DRS and to our color corrected least-squares deconvolution profiles.}
	\label{fig:fwhmlsdK}
	\end{figure}

Then, for each LSD profile in absorption, we produce a normalised positively defined probability distribution function by subtracting their residual continuum (see Sec.~4.2.2 on \citealt{berdinas2016}) and normalising its area to unity. Finally, we calculate the FWHM-LSD index as the full-width-at-half-maximum of a Gaussian function fitted to the mean-line profiles. The FWHM-LSD is a proxy of the changes of the profile width. In Figure~\ref{fig:fwhmlsdK} we show how, in contrast to the FWHM of the CCFs (FWHM-CCFs), the FWHM-LSD measurements do not correlate with the changing slope of the observed SEDs. That is, with the index $\kappa$, that accounts for the  SED slope change in linear region (see blue area in Figure~\ref{fig:sed}). Both indices, the FWHM-CCF and the FWHM-LSD resulted to have different mean values. The side lobes of the CCFs may necessarily derive in an underestimation of the continuum, and so of the FWHM of the Gaussian fitted to the CCFs by the DRS.

Uncertainties in the LSD profiles, and thus in the FWHM-LSDs are difficult to estimate analytically. Instead, we used an
empirical procedure based on the fact that uncertainties follow the SNR of the observations. We assumed the standard deviation
of the differences between FWHM-LSD datapoints to be $\sqrt{2}$ times the uncertainty of the mean SNR of each night (at
reference \'echelle aperture number 60). We used as reference the first observation of each night, then the error values of the observations within the night were obtained by scaling this standard deviation by a factor of \textless SNR\textgreater/$\mathrm{SNR}_{\mathrm{obs}}$. 

\begin{figure}
	\centering
	\includegraphics[width=\columnwidth]{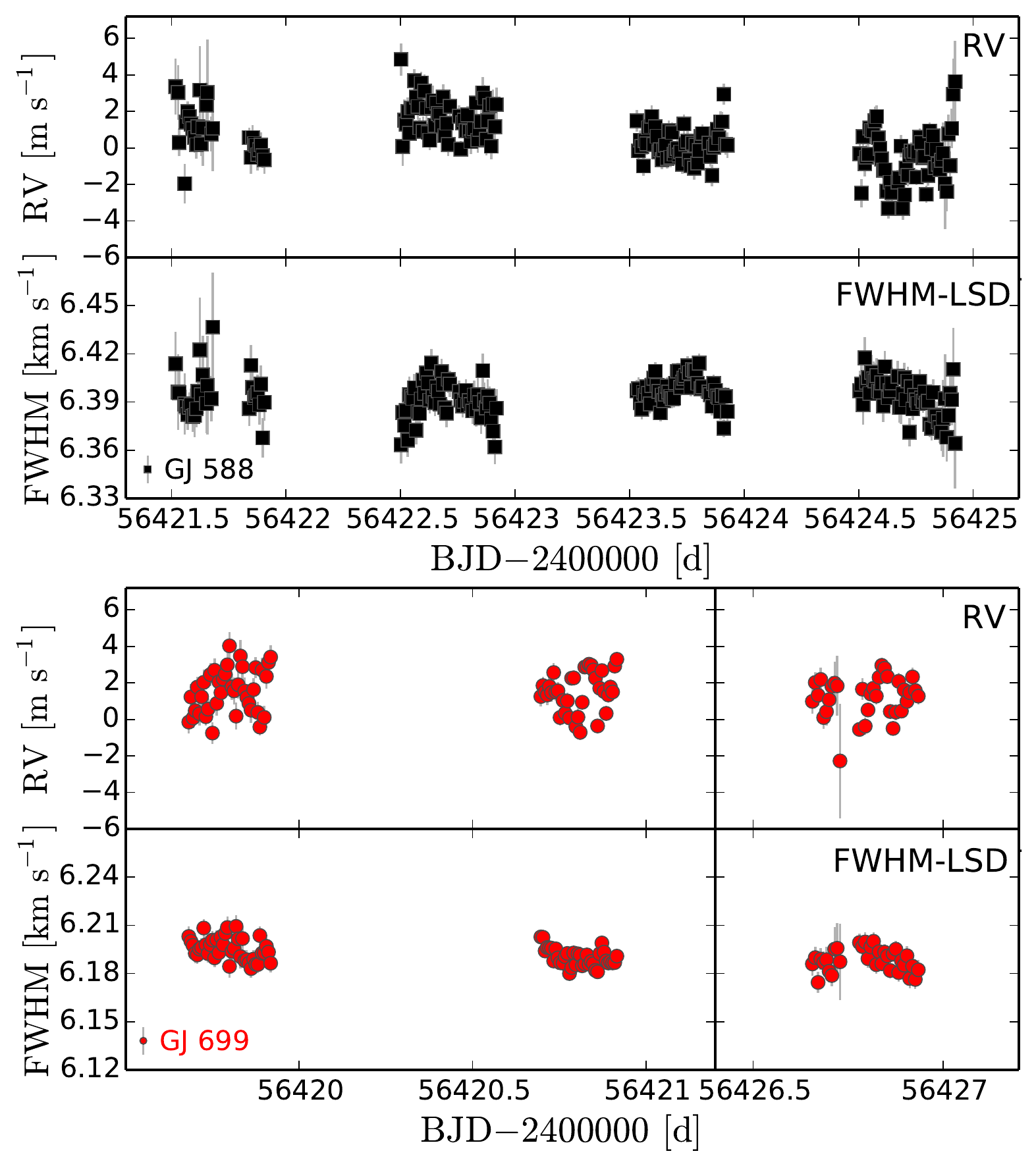}
	\caption{RV and FWHM-LSD time-series used in this study and resulting after applying several intra-night and night-to-night corrections. The upper and bottom panels show the four and three high-cadence observing nights of GJ~588 (black squares) and GJ~699 (red dots), respectively. The last GJ~699 observed night was observed four nights after the second.}
	\label{fig:timeser}
	\end{figure}

The final measurements to be used for this study were the RVs and the FWHM-LSDs corrected as explained above. See in Table~\ref{tab:data}  the original and corrected data (full version included in the online material). In Figure~\ref{fig:timeser} we show the final resulting time-series. The FWHM-LSD still shows some patterns which seem to repeat themselves on different nights (e.g. the arc-shape of the FWHM-LSD in nights two and three of GJ~588). Since we have corrected the spectra from the SED normalisation effects,
such variability can not be driven by flux distribution changes. An explanation could come from the barycentric broadening of
the spectral lines caused by changes of the star-Earth differential velocity during the observations. That is, by changes in
the star-Earth relative RV between observations taken close and far from the zenith\footnote{Bodies at the zenith seem to move faster for a fixed observing time causing a broadening of the spectral lines.}. However, our measurements indicate that the contribution of such effect is negligible at the HARPS precision level for these targets. Nevertheless, even when we cannot identify the origin, we know that the periods associated to such pattern will be close to 1-day and/or sub-multiples in the periodograms. Therefore, our study ranging from 20~min to 3~h precludes any misleading signals coming from this pattern.

\begin{table*}
\begin{center}
\caption{GJ~588 and GJ~699 RV and FWHM data used in this study. The columns are: the star, the barycentric Julian date (BJD), the observed radial velocities obtained with TERRA (RV), the observed full-width-at-half-maximum index obtained from the cross-correlation function by the HARPS pipeline (FWHM-CCF), the radial velocities corrected from the instrumental effects as described in Section~\ref{sec:systematics} (RVc), and the full-width-at-half-maximum corrected from the SED normalisation effect (FWHM-LSD). This study analyses the data of the last two columns.}     
\label{tab:data}
\begin{tabular}{cccccccc}\hline
OBJECT  & BJD & & RV & FWHM-CCF & & RVc & FWHM-LSD \\
            & days & &$\rm m\,s^{-1}$ & $\rm m\,s^{-1}$ & & $\rm m\,s^{-1}$ & $\rm m\,s^{-1}$\\ \hline         
GJ~588 & 56421.518 & & $5.55\pm1.61$ & $3046.74\pm3.87$ & & $3.36\pm1.52$ & $6413.88\pm19.79$ \\
``	  & 56421.528 & & $5.30\pm1.66$ & $3046.01\pm4.91$ & & $3.04\pm1.50$ & $6396.19\pm23.49$ \\
``	  & 56421.533 & & $2.36\pm0.88$ & $3049.73\pm2.07$ & & $0.30\pm0.76$ & $6395.80\pm13.58$ \\
``	  & ... & & ... & ... & & ... & ...\\
GJ699 & 56419.682 & & $0.00\pm0.75$ & $2981.66\pm1.77$ & & $-0.15\pm0.61$ & $6203.06\pm 6.08$\\
``	 & 56419.688 & & $1.42\pm0.86$ & $2982.55\pm1.75$ & & $1.23\pm0.72$ & $6200.08\pm6.06$\\
``	 & 56419.694 & & $-0.24\pm0.78$ & $2984.66\pm1.78$ & & $0.07\pm0.56$ & $6197.53\pm6.12$\\ 
``	  & ... & & ... & ... & & ... & ...\\\hline
\end{tabular}   
\end{center}
This table is available in its entirety in a machine-readable form in the online journal. A portion is shown here for guidance regarding its form and content.
\end{table*}

\section{Analysis}\label{sec:analysis}

\subsection{Short time domain variability}\label{sec:periodicsignals}

We used periodograms to search for periodic signals embedded in the RV and FWHM-LSD high-cadence time-series. Periodograms are
plots which represent a reference statistic in the y-axis versus a range of  periods in the x-axis. This reference statistic accounts for the improvement of fitting the data with a sinusoidal model compared to an initial hypothesis (e.g. no periodic signals). Classic Lomb-Scargle periodograms \citep{lomb1976, scargle1982} use the F-ratio reference statistic; instead, we used the difference of the logarithm of the likelihood function
as the reference statistic \citep[\deltaL\ periodograms, see][]{baluev2009}. Compared with the Lomb-Scargle periodograms, the likelihood periodograms have several advantages. The most important one is that the likelihood function allows a \emph{global} search, in the sense that all parameters (including noise parameters such as the ``stellar jitter'') are optimized at the period search level. This is specially important in the case of M dwarfs, which are all affected by activity to a certain extent.

For a grid of periods uniformly sampled in the frequency space our likelihood periodograms search the model best fitting the data. We defined this model as:
\begin{equation}
	v_{i,\mathrm{Night}} = \gamma_{\mathrm{Night}} + \dot{\gamma}(\Delta t_i) + \mathcal{K}_p(\Delta t_i)\,\,\,,
	\end{equation} 
where $\gamma_{\mathrm{Night}}$ and $\dot{\gamma}(\Delta t_i)$ are parameters accounting for an offset velocity and a linear trend, respectively. Note that we let each night have a different velocity offset, whereas a single linear trend is fitted to all nights. This means we have modeled the Doppler time-series considering each high-cadence night as an independent set. The analysis procedure is equivalent to that used in \citealt{anglada2016a}, where we considered different instruments as independent datasets. We proceeded in that way aiming at filtering out any possible periodicity longer than 1 day in which we were not interested when searching for M dwarfs stellar pulsations. The 1-day signals typically populate periodograms and,  even when we have made an effort to correct our data from nightly jumps (see Section~\ref{sec:systematics}), this is also the case here. Most probably other unknown intra-night systematic effects drift the RVs during the observations causing these 1-day periods and/or sub-multiples. In other words, our approach prevents a superposition of $p$-th sinusoids that otherwise would be needed in the model before we were able to study sinusoids compatible with the pulsation range of interest. As a comparison, we show in Appendix~\ref{app:A} the result when all nights are analysed as a single dataset. On the other hand, the linear trend parameter allows to account for possible long trends caused by a long term acceleration. Finally, $\mathcal{K}_p(\Delta t_i)$ is a sum of $k$ sinusoids and can be written as:
	\begin{equation}
	\mathcal{K}_p(\Delta t_i)=\sum_p^k{A_p}\sin{\left(\frac{2\pi}{P_p}t_i\right)}+B_p\cos{\left(\frac{2\pi}{P_p}t_i\right)}\,\,\,,
	\end{equation} 
where each $p$-th sinusoid is defined by the $A_p$ and $B_p$ amplitudes and the period $P_p$. As we said, the likelihood periodograms allow a \emph{global} search, and that means we do not apply pre-whitening procedures. Instead, every time that a model with $p$+1 sinusoids (SOLp+1) is preferred over the $p$ case (SOLp) a new sinusoid is added to the model. Then, all the model parameters are re-adjusted every time that we search for those with the new added sinusoid. With this model we are assuming that stellar pulsations can be modeled with sinusoidal functions. This assumption can be untrue, but this is the simpler model we can apply in the absence of information about the real nature of the M dwarfs stellar pulsation signals. Moreover, similar models have demonstrated to be useful in the search of pulsations in other spectral types. For more details about the model see \cite{anglada2016a}.

\begin{figure}
	\centering
	\includegraphics[width=\columnwidth]{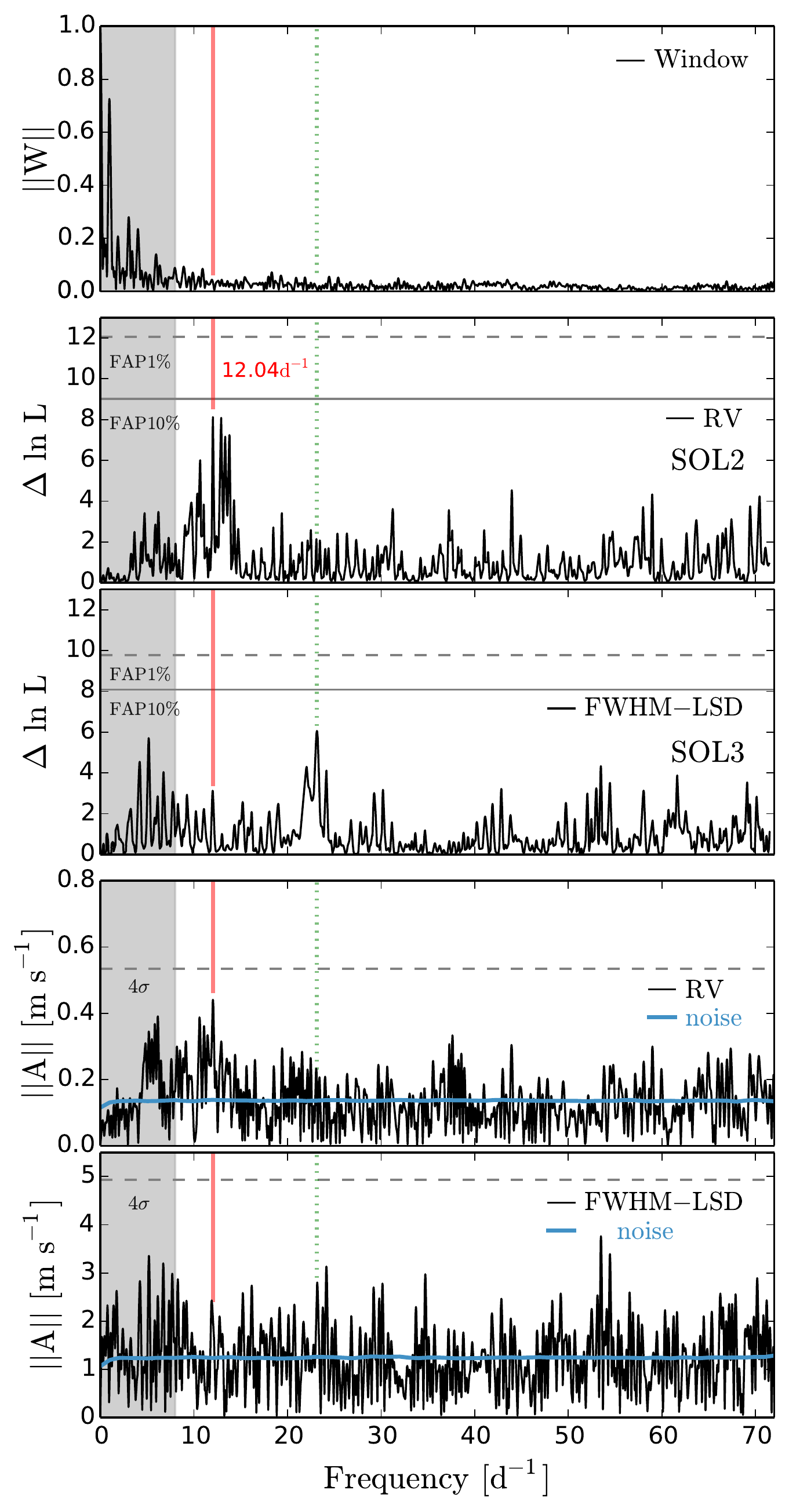}
	\caption{GJ~588 analysis of periodicities. GJ~588 window function (first panel), RVs and FWHM-LSD likelihood periodograms (second and third panels), and RVs and FWHM-LSD power spectra (fourth and fifth panels). The grey areas indicate frequencies out of the predicted pulsational range. The vertical red line highlights a putative signal at $12.04\,\rm{d}^{-1}$ appearing in both the RVs likelihood periodogram and power spectra. The vertical dotted green line highlights a peak found in the RVs of GJ~699 for which we detect a counterpart in the FWHM-LSDs of GJ~588 (third panel). Horizontal lines account for different levels of significance. The blue lines in last two panels indicate the noise power spectra. See main text for details.}
	\label{fig:periodograms}
	\end{figure}

\begin{figure}
	\centering
	\includegraphics[width=\columnwidth]{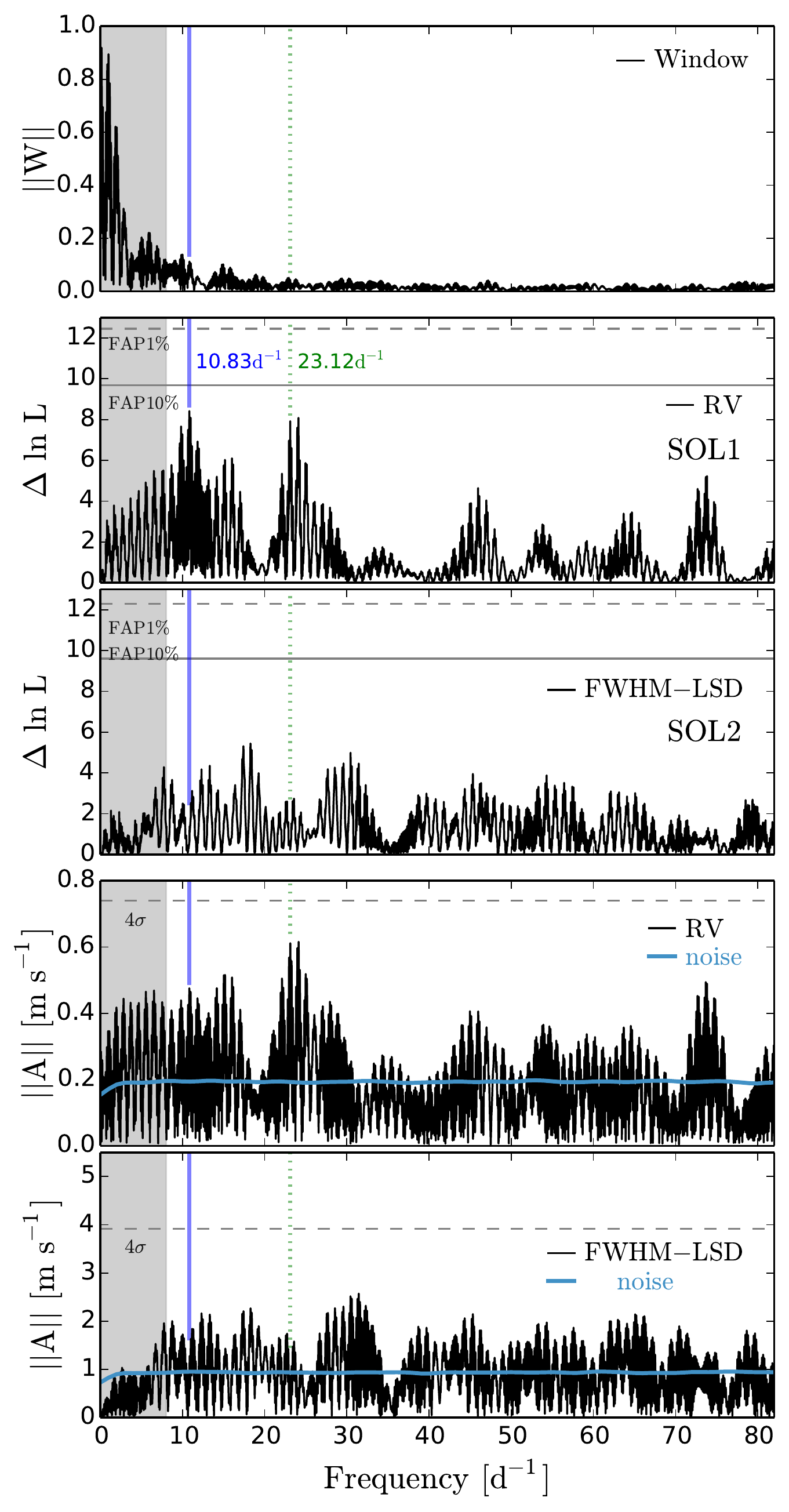}
	\caption{GJ~699 analysis of periodicities. Panels follow the same criteria as in Fig~\ref{fig:periodograms}. The interruption of the continuous consecutive night monitoring complicates the window function causing aliases. Like for GJ~588, no significant signals were detected, but some non-negligible structure is present. The two preferred peaks are at 10.83 (vertical blue line) and at $23.12\,\rm{d}^{-1}$ (vertical dotted green line), also present in the FWHM-LSD periodogram of GJ~588.}
	\label{fig:periodograms699}
	\end{figure}

The second and third panels of Figure~\ref{fig:periodograms} show the RV and FWHM-LSD likelihood periodograms of GJ~588. The same panels of Figure~\ref{fig:periodograms699} show the likelihood periodograms for GJ~699. Grey areas in these panels highlight the range of frequencies where pulsations are not expected (i.e. periods out of the 8-$72\,\rm{d}^{-1}$ frequency range). In the upper right corner of these panels we indicated the number of $p$ sinusoids included in the model until the first solution outside the grey area is found. For example, the SOL2 in the upper right corner of the second panel of Figure~\ref{fig:periodograms} means the model fitting the RVs of GJ~588 contains two sinusoids, the first one corresponding to a large amplitude signal with a period out of the 20~min--3~h range. In particular, the periods of these sinusoids were: in the case of the RVs of GJ~588 $\sim0.5$~days (SOL1),   for FWHM-LSDs of GJ~588 $\sim1$ and $\sim0.4$~days (SOL1 and SOL2) , and $\sim0.3$~days (SOL1) in the case of the FWHM-LSD of GJ~699.

\begin{figure}
	\centering
	\includegraphics[width=\columnwidth]{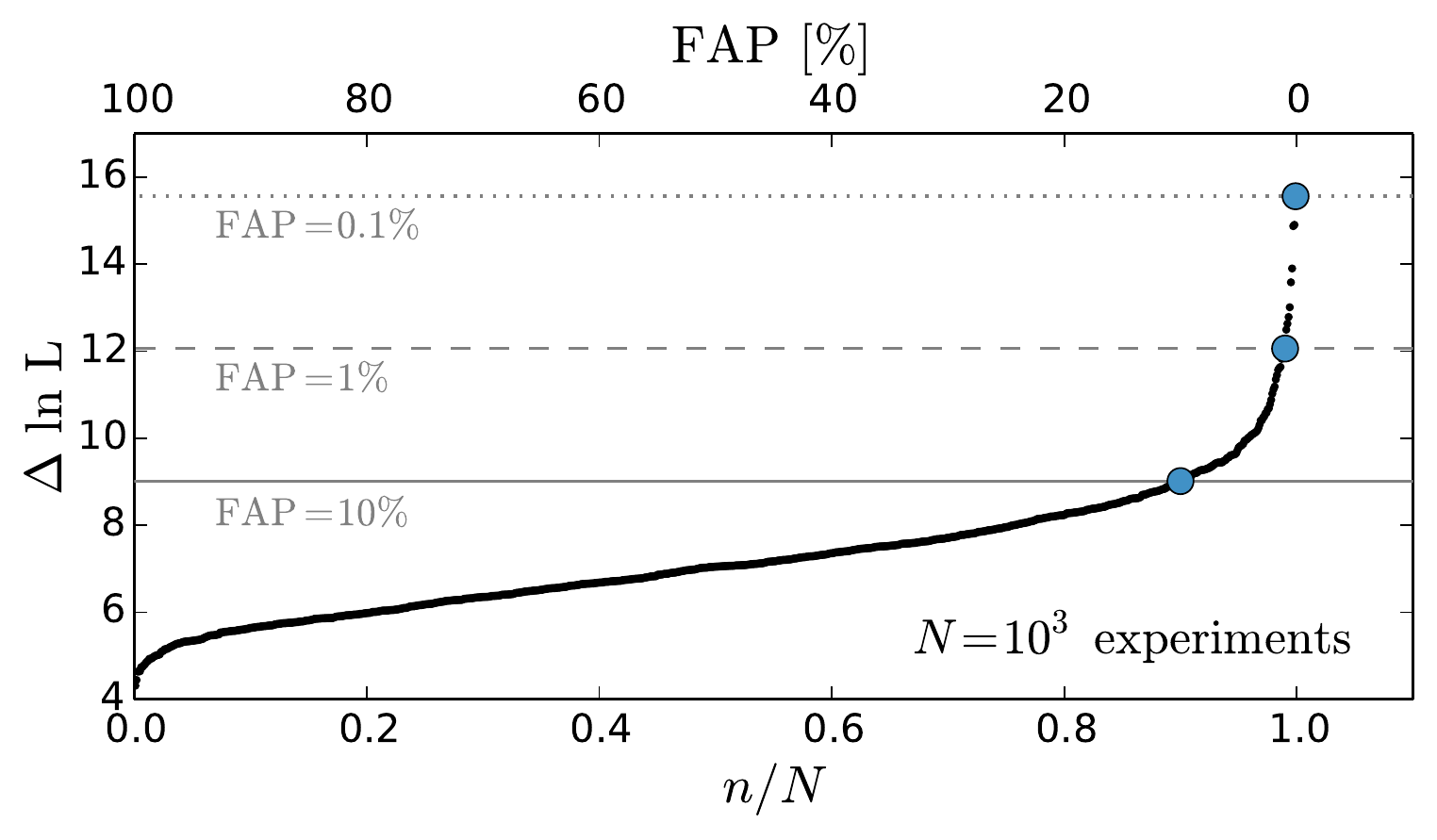}
	\caption{Bootstrapping approach used to calculate the FAP thresholds. The y-axis corresponds to the maximum \deltaL\ values resulting for each of the N=$10^3$ bootstrapping experiments we performed. The x-axis accounts for $n/N$, the cumulative probability of obtaining a certain \deltaL\ (i.e. the \deltaL\ values are sorted and then plotted versus $n/N$). The highest \deltaL\ obtained corresponds to the lowest false alarm probability (FAP=0\%) and the other values are obtained from this assumption. Horizontal lines and blue dots indicate the 0.1\%- 1\%- and 10\%-FAP thresholds. This experiment corresponds to the RVs of GJ588, i.e. these are the FAP thresholds plotted on the second panel of Figure~\ref{fig:periodograms}.}
	\label{fig:fap}
	\end{figure}

As a figure of merit to quantify the significance of a detection we used the so-called `False Alarm Probability'' (FAP
hereafter), which accounts for the probability of obtaining a peak by a random combination of the noise \citep{cumming2004}. A FAP=1\% is considered the minimum threshold that a peak has to reach to claim it as a detection \citep{cumming2004,baluev2009}. The grey dashed lines in the likelihood periodograms of Figures~\ref{fig:periodograms} and \ref{fig:periodograms699} highlight the 1\%-FAP thresholds. As a comparison, the 10\%-FAP is also shown with a solid line. The FAP can be calculated analytically, but it is required the number of independent frequencies in the range of study. Given that this number is ambiguous when the sampling is not regular, we estimated the FAP empirically. The goal was to establish how the maximum likelihood statistic is distributed in the presence of noise only. To do this, we generated synthetic data by randomly permuting the measurements among the given observed epochs (i.e we bootstrapped the time-series). Then we calculated periodograms over the
resulting series and we recorded the maximum \deltaL\ achieved in each of these test periodograms. The FAP of a signal is the number of synthetic
experiments giving spurious \deltaL\ larger than the original time-series divided by the total number of tests. Thus, after repeating this experiment $N=10^3$ times, the highest \deltaL\ obtained was made to correspond to a 0\%-FAP. From this assumption we can easily derive the 0.1\%-, 1\%-, and 10\%-FAP thresholds as we show in Figure~\ref{fig:fap}. That is, plotting the maxima \deltaL\ sorted increasing order versus $n/N$, the cumulative number of experiments, and interpolating the result.
	
No signals were found with a FAP $<$1\% in the range of frequencies where we expected to find pulsations. However, for the RVs,
we can see some excess of power for both targets, GJ~588 and GJ~699, in the range of a few hours (or tens of d$^{-1}$) close to the
10\%~FAP. In particular, the preferred peaks have a corresponding frequency of $12.04\,\rm{d}^{-1}$ (indicated with a vertical
solid red line in Fig.~\ref{fig:periodograms}) in the case of GJ~588, and of 10.83 and $23.12\,\rm{d}^{-1}$ (vertical solid blue
and dotted green lines in Fig.~\ref{fig:periodograms699}) in the case of 
GJ~699. Nevertheless, the presence of the
$23.12\,\rm{d}^{-1}$ peak also in the FWHM-LSD periodogram of GJ~588 (highlighted in Fig.~\ref{fig:periodograms} also with a
vertical dotted green line) lead us to suspect that it could be an instrumental artifact. The $12.04\,\rm{d}^{-1}$ is recovered
regardless of the approach used to filter out the periodicities close to 1-day (and their aliases). However, we decided to
consider each night as independent datasets because in this case the model requires fewer sinusoids \footnote{For example, in the case of GJ~588 RVs if we consider each night independently only one sinusoid is required to account for the long-term variability (SOL1 corresponds to $\sim0.5$~days). On the contrary, if we consider the run as a whole, the $12.04\,\rm{d}^{-1}$ period is recovered after fitting two $\sim$1.5 and $\sim$0.5~days sinusoids. See Appendix~\ref{app:A} for more details.}. The FWHM-LSD periodograms do not present any other peak besides the suspected spurious at $23.12\,\rm{d}^{-1}$ for GJ~588, and a peak at $53.5\,\rm{d}^{-1}$ that we think is an artefact since it arises in the power spectra but not in the likelihood periodogram. But, even when some structure is present, any peak is not even close to the permissive threshold of 10\%~FAP, for either GJ~588 or GJ~699.

In order to perform an independent assessment of the possible signals, we have also computed the power spectra using the code SigSpec
\citep{reegen2007}. This code uses a pre-whitening methodology, i.e. it does not globally model all the solutions. As a consequence, in this case we used as input the time-series of the residuals to the best models found with the likelihood
periodograms, so as not to introduce any undesirable trend as a result of the pre-whitening process of signals in the day or sub-day range (see \citealt{anglada2015a} for more details). Thus, we used the residuals to SOL1 ($\sim0.5$~days) for the RVs and to SOL2 ($\sim1$ and $\sim0.4$~days) for the FWHM-LSD of GJ~588; and in the case of GJ~699, the
original RVs and the residuals to SOL1 ($\sim0.3$~days) for the FWHM-LSD. The power spectra relies on the discrete Fourier transform (DFT) and
consists of plotting its amplitude (or the square root of the sum of its real and imaginary squared components) versus a grid
of frequencies. The last two panels of Figures~\ref{fig:periodograms} and ~\ref{fig:periodograms699} show the RV and
FWHM-LSD power spectra. The general criterion to consider a peak statistically significant in a power spectra is to reach  at least four times the signal/noise amplitude ratio \cite{breger1993}. This rule is the commonly known as the ``4~$\sigma$ criterion''. Note that, besides the name, this concept is not directly the probability of a normal distribution. Statistical conclusions that attach to regular spaced time-series do not apply to non-equally sampled data as is the case of our observations. The grey dashed lines in the last two plots indicates the 4~$\sigma$ threshold. It was calculated as four times the mean amplitude of the
peaks with frequencies higher than $8\,\rm{d}^{-1}$ (white area, pulsation range). Again, even when no peak reaches the
threshold, we obtained a power excess in the range of a few hours with a preferred peak at $12.04\,\rm{d}^{-1}$ in the RVs of GJ~588. In
the same way, the power spectra of GJ~699 showed results comparable to those of the likelihood periodograms. 

We also calculated respective window functions (first panel of Figures~\ref{fig:periodograms} and \ref{fig:periodograms699}) using a
DFT. This function helps to identify misleading peaks arising from periodicities caused only by the sampling. The reason for the more complicated window function for GJ~699 is the time gap between the second and the third observing night which causes more aliases. However, since
the resolution of the individual peaks is inverse to the total time baseline, this is also the reason why in the GJ~699
periodogram the peaks are narrower. Additionally, the window function is a very useful tool because any peak corresponding to a real
periodic signal in the time domain, rather than being a simple Dirac Delta function in the power spectra, is a convolution of it
with the window function \citep{gray1973}. In this case, the window function also indicates that the putative signal at $12.04\,\rm{d}^{-1}$ in
the case of GJ~588 is far away from the range of influence of the window peaks.

Aiming at checking if the sampling could originate an excess of power in a certain zone, we have also generated the power spectra of the noise. That is, we calculated the averaged power spectra resulting from $10^3$ bootstrapping experiments. Such power spectra were found to be almost flat, indicated by the horizontal blue lines of the bottom panels of Figures~\ref{fig:periodograms} and \ref{fig:periodograms699}. This result means that the sampling does not bias the spectrum of the noise in the absence of signals. Therefore, even when the $12.04\,\rm{d}^{-1}$ or the 10.83 and $23.12\,\rm{d}^{-1}$ peaks do not
reach our threshold criterion, the sampling seems not to be the cause. 

\subsection{Compatibility with pulsation models}\label{sec:putativesignal}
The discussed putative signals may be caused by intrinsic oscillations of the star that might not be necessarily described as single
frequency sinusoids. Therefore, as a first step, we have calculated where we could expect to find pulsations with these
putative periods in the Hertzsprung-Russell (HR) diagram. This approach will allow us to know if excited models are compatible with the star parameters, i.e. if the star and any excited model with similar physical parameters share the same position in the HR diagram. With this aim, we used the evolutionary tracks from RL14, which were calculated from different models with a range of masses of 0.10-0.60~M$_\odot$ with 0.05 resolution (see more details about these models in table~1 from RL14). Later, we perturbed models along these tracks and we explored their pulsation instabilities for modes of degree $\ell=0$ to 3. 

\begin{figure}
    \includegraphics[width=\columnwidth]{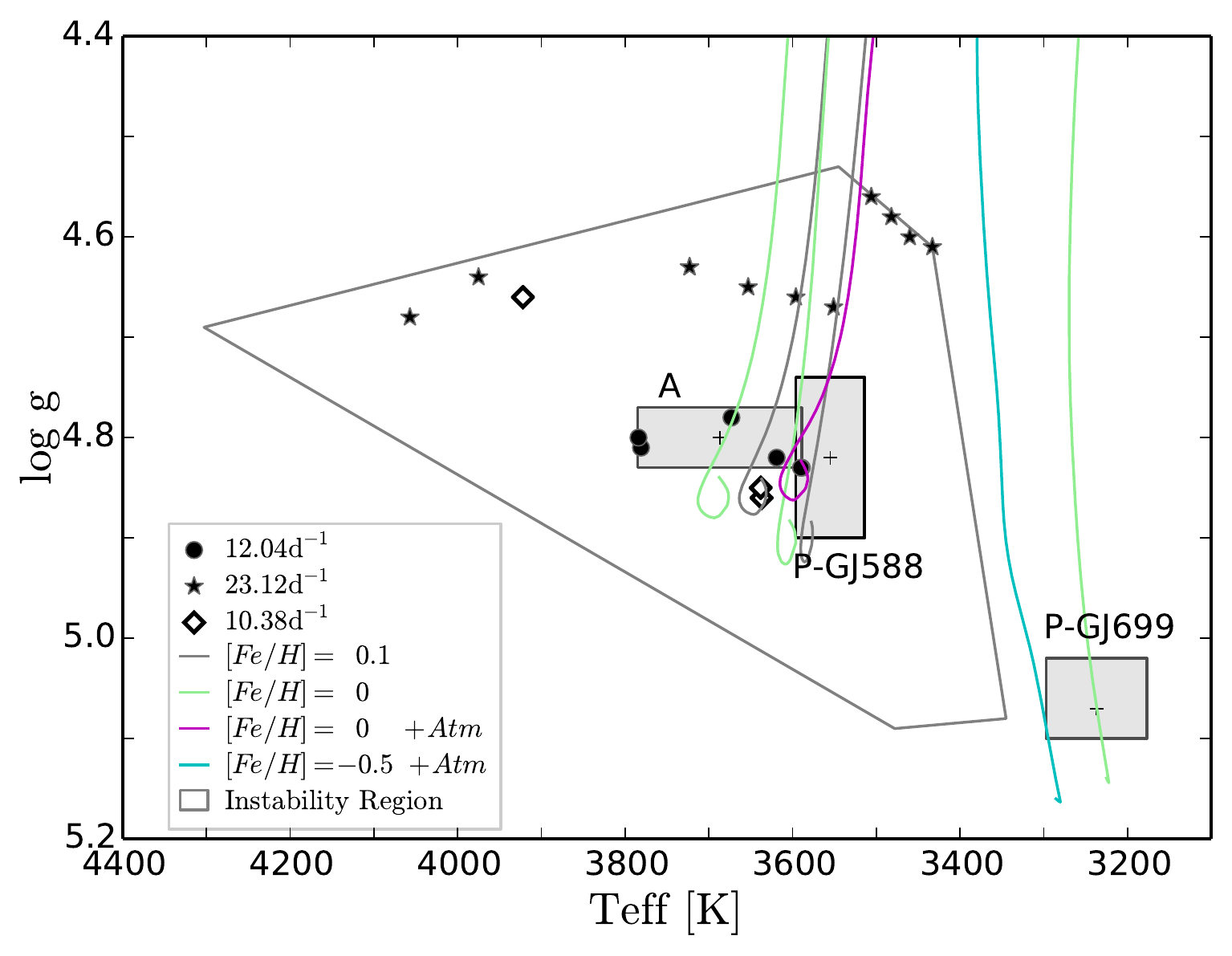}
    \caption{Photometric (P--GJ~588 and P--GJ~699) and asteroseismic (A) boxes in the T$_\mathrm{eff}$-$\log g$ diagram. We only plotted the evolutionary tracks in RL14 that cross any of these boxes and have metallicities and masses closer to the literature values (0.40 and 0.45~M$\odot$ and super-solar metallicity for GJ~588, and 0.15~M$\odot$ and sub-solar for GJ~699). Solar metallicity was also included for comparison purposes. Stellar models including an atmosphere are indicated with the label ``Atm'' in the legend. All tracks were calculated with a mixing length parameter $\alpha=1$. Solid black dots indicate models excited with a 12.04~d$^{-1}$ period (putative signal of GJ~588). Stars and diamonds correspond to models with $23.12\mathrm{d}^{-1}$ and $10.38\mathrm{d}^{-1}$ periods (putative signals of GJ~699). The black dot in the intersection of the A and P--GJ~588 boxes (0.45~M$_\odot$ on the magenta track) give theoretical support to the GJ~588 putative signal.}
    \label{fig:tracks}
    \end{figure}

In Figure~\ref{fig:tracks} solid black dots show the stellar models excited with periods in the range of the
observed one for GJ~588  ($\sim$12~d$^{-1}$). The physical parameters of those excited models set up the ``asteroseismic box''
of GJ~588 (see A box in Fig.~\ref{fig:tracks} defined by $T_\mathrm{eff}=3687\pm98$~K and $\log{g}=4.80\pm0.03$). The
``photometric box'' of GJ~588 was defined to assess if the physical parameters of GJ~588 are compatible with those of the models
comprised in the ``asteroseismic box''. We used the $T_\mathrm{eff}$ and $\log{g}$ values given in the literature to define the
box. In particular, we used T$_\mathrm{eff}$=3555$\pm$41~K, and mass and radius determinations of 0.43$\pm$0.05~M$_\odot$ and
0.42$\pm$0.03~R$_\odot$ from \cite{gaidos2014} to derived a $\log g$=4.82$\pm$0.08 (see Table~\ref{tab:params})\footnote{\cite{neves2013} and \cite{bonfils2013} give $T_{\mathrm{eff}} = 3325\pm80\,\rm K$ for GJ~588. However, from the evolutionary tracks in \cite{baraffe1998} and RL14 this value is in clear contradiction with the mass given by the same authors: thus, if $M=0.47\,\rm M_{\odot}$ holds, then $T_{\mathrm{eff}}>3400\,\rm K$; otherwise the mass would correspond to $\sim0.2\mathrm{M}_{\odot}$ models.}.

Additionally, in Figure~\ref{fig:tracks} we show the ``photometric box'' of GJ~699 defined by T$_\mathrm{eff}$=3237$\pm$60 \citep{gaidos2014} and $\log{g}=5.040\pm0.005$ derived from 0.15$\pm$0.02~M$_\odot$ and 0.187$\pm$0.001~R$_\odot$ \citep{boyajian2012}, along with the black stars pointing the models that excite periods at $\sim$10.83~d$^{-1}$ and diamonds for periods at 23.12~d$^{-1}$. We also show the evolutionary tracks that cross the photometric or the asteroseismic boxes of GJ~588 and GJ~699. For the sake of clarity, we only show the evolutionary tracks that have similar metallicity and mass as the stars under study.

In the case of GJ~588, the ``photometric box'' resulted to be traversed by twelve tracks with masses between 0.15 and
0.50~M$_\odot$; while the ``asteroseismic box'' was traversed by 19 tracks with masses in the 0.20 to 0.50~M$_\odot$ range.
Consequently, evolutionary tracks within the boxes encompass the 0.43~M$_\odot$ mass determination from the literature. The
excited models within the GJ~588 asteroseismic box have masses in the 0.45 to 0.50~M$_\odot$ range corresponding to low-radial, low
degree, $\ell$=1, and $\ell$=2, {\em g}-modes. In particular, the excited model in the overlapping boxes has 0.45~M$_\odot$,
which would give theoretical support to the putative signal.

On the contrary, results for GJ~699 point to a very different situation. In this case, even when the tracks falling within the
``photometric box'' are in good agreement with the mass determination in the literature (6 tracks with 0.10--0.20~M$_\odot$). The excited models at $\sim$23.12~d$^{-1}$ ($\sim$63~min) and $\sim$10.38~d$^{-1}$ ($\sim$2.3~h), indicated in Figure~\ref{fig:tracks} with stars and diamonds respectively, correspond to masses (0.40--0.60~M$\odot$) that do not comprise GJ~699 lower mass determination from literature. Therefore, our pulsation analysis does not support the presence of oscillations in GJ~699, which we recall was out (but close to the edge) of the instability region.

\subsection{Completeness and signal detectability limit in the sample}\label{sec:detlim}

In this section, we set up an upper limit for the intra-night precision of the HARPS spectrograph. Therefore, if stellar pulsations
on M dwarf stars exist and induce Doppler shifts in the spectra, such upper limit would indicate the amplitude threshold that
we would be able to detect with a CTB-like campaign in either the RVs or any other index.

We performed the following experiment using GJ~588 as reference. We preferred GJ~588 over GJ~699 because, even when they are
equally stable, GJ~588 has more datapoints, and a simpler window function. Firstly, we randomized the RV measurements to create
the time-series of the noise with the observed time-span. Later, we added simulated sinusoids of increasing amplitudes (from
0.2 up to $0.8\,\mathrm{m\,s}^{-1}$ in steps of $0.02\,\mathrm{m\,s}^{-1}$) and random phases. Besides, we tested different
frequencies ranging from 0 to $70\,\mathrm{d}^{-1}$ to sample the pulsation frequency domain and to measure the dependence of
the threshold with frequency. Secondly, we calculated likelihood periodograms and we checked when our method succeeded in recovering
the amplitude and frequency injected (``positive experiment''). In particular, we performed one hundred experiments for each
input amplitude and frequency, varying the noise in all cases. The criterion we chose to define an experiment as positive was
to recover both an amplitude within $\pm\,40\%$ of the input value and a frequency within the resolution range given by the
inverse of T, the total time-span of the observing run. Such criteria accounts for the typical large uncertainties of the amplitudes close to the detection limit and ensures the rejection of the experiments with large FAPs. 
 
\begin{figure}
	\centering
	\includegraphics[width=\columnwidth]{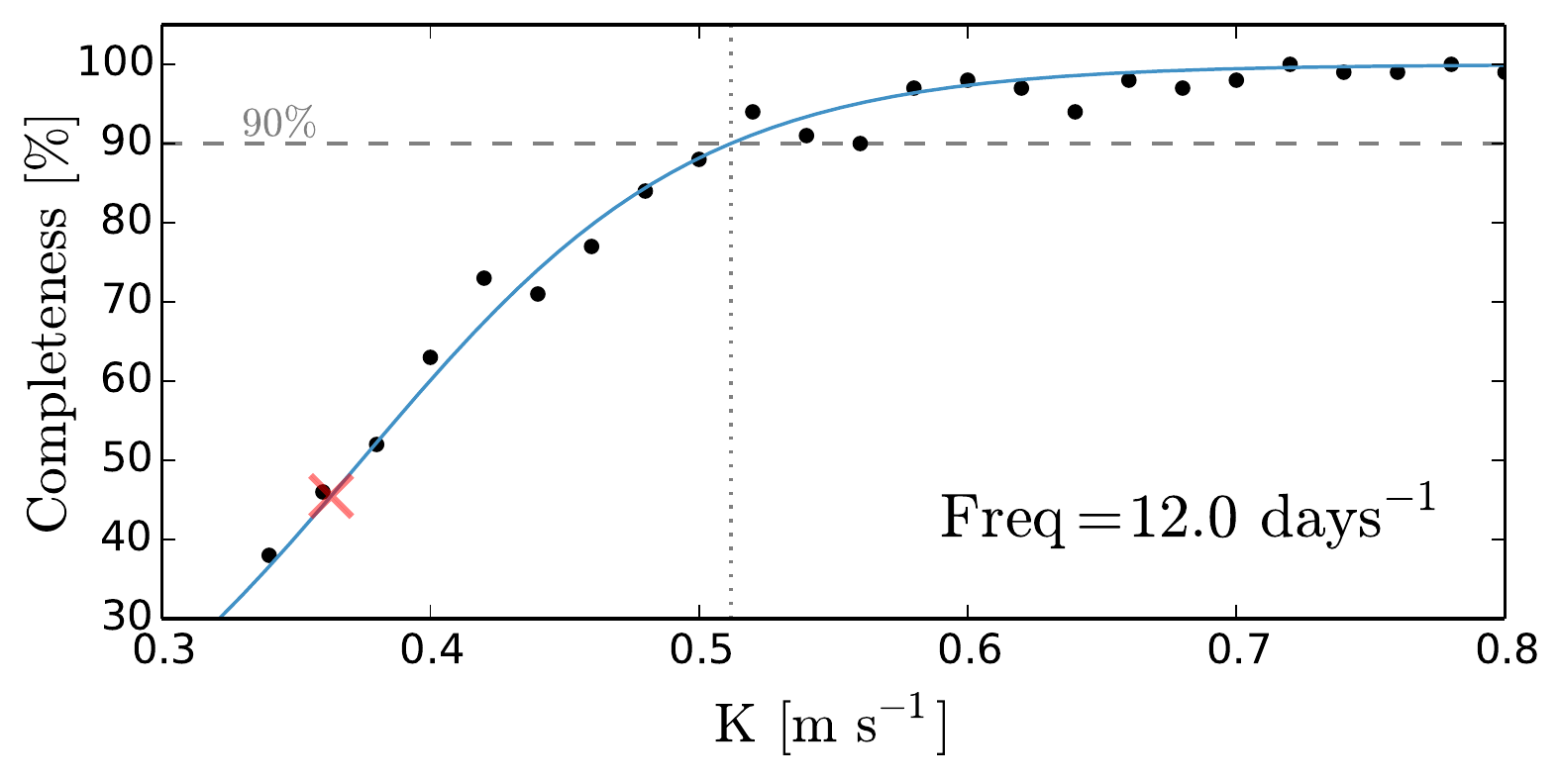}
	\caption{Completeness assessment of a $12\,\mathrm{d}^{-1}$ putative signal. The black dots indicate the proportion of experiments (completeness, y-axis) for which our analysis tools can recover the true amplitude (K, x-axis) injected at the different simulated input sinusoids. The blue solid line is an S-shape function fitted to the black dots. The horizontal grey dashed line indicates the 90\% completeness level. A periodic physical phenomenon with a frequency of $12.0\,\mathrm{d}^{-1}$ should produce a Doppler signal with a minimum amplitude of $0.51\,\mathrm{m\,s}^{-1}$ (vertical grey dotted line) to be easily detected with HARPS with a 90\% probability. The red cross ($0.36\,\mathrm{m\,s}^{-1}$) indicates the amplitude of the putative signal detected on GJ~588. It corresponds to a low 45\% completeness.}
	\label{fig:conflevel}
	\end{figure}

Finally, we defined the completeness at a certain frequency as the
percentage of positive experiments recovered at each input amplitude; thus, the completeness increases with increasing
amplitude, as expected. Consequently, fitting a simple S-shape function (or sigmoid function) we could define the limiting
amplitude required to reach a 90\% completeness with HARPS (see Figure~\ref{fig:conflevel}).

\begin{figure}
	\centering
	\includegraphics[width=\columnwidth]{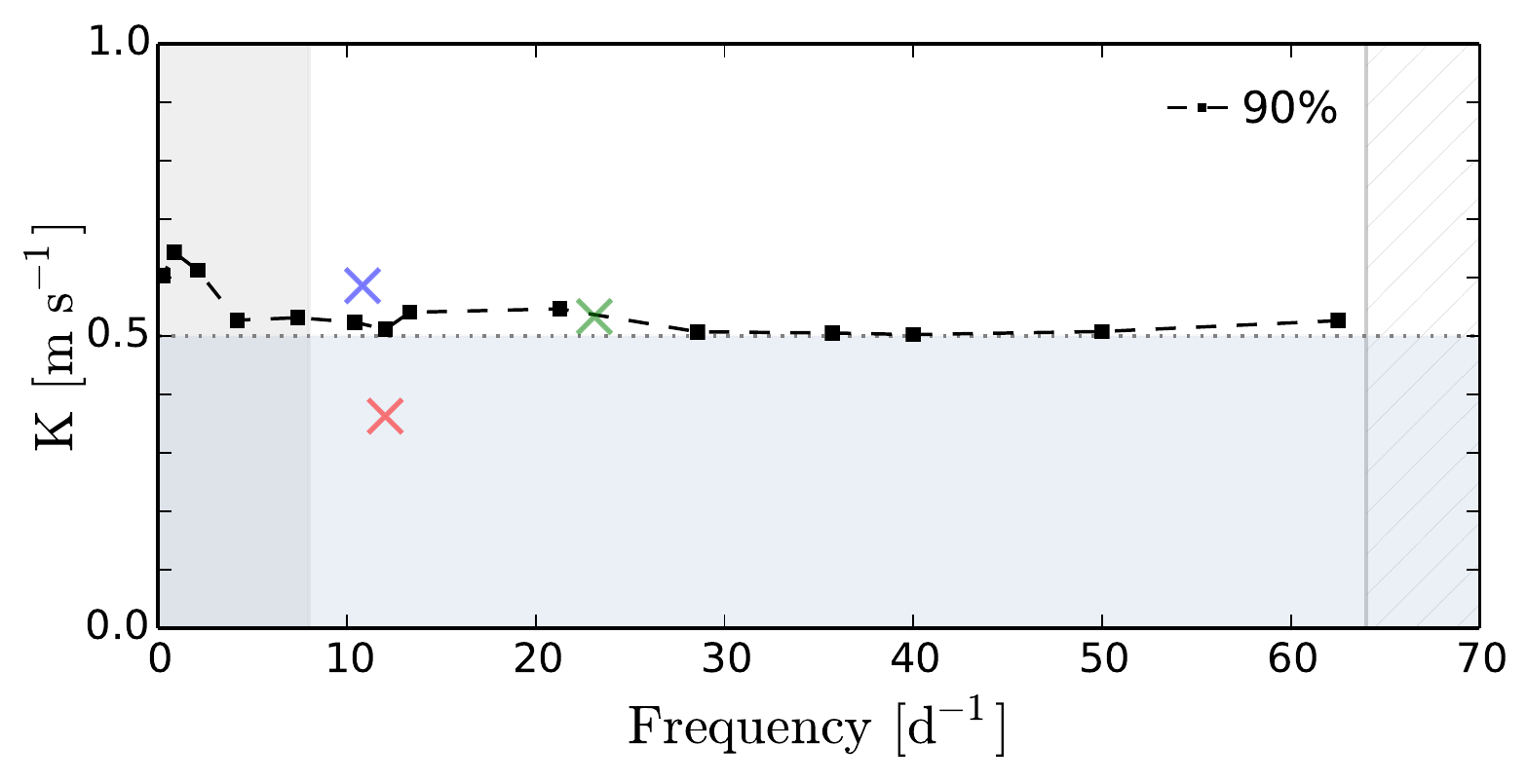}
	\caption{Limiting amplitudes detectable with a 90\% completeness as a function of the frequency (black dashed line). The grey left area is out of the pulsation range. The stripped right area indicates frequencies non-accessible for being of the order of the exposure time of the observations. The blue area highlights the general range of inaccessible signals, i.e. those with amplitudes below $0.5\,\rm m\,s^{-1}$. The GJ~588 putative signal (red cross) is below the amplitude detection limit. The GJ~699 putative signals (blue and green crosses) are above the limit but the poorer sampling of the GJ~699 observations degrade the completeness 90\% limit up to
$\sim1.20\,\mathrm{m\,s}^{-1}$ seriously diminishing our sensitivity.}
	\label{fig:detlim}
	\end{figure}

Results for different input frequencies within the stellar pulsation range shown in Figure~\ref{fig:detlim} indicate that, in
spite of the frequency dependence, if stellar pulsations exist and induce Doppler signals, we will not be able to detect them
with HARPS if the induced signal has an amplitude below $0.5\,\mathrm{m\,s}^{-1}$. The putative signal found in the GJ~588 RVs at
$12.04\,\mathrm{d}^{-1}$ (red solid vertical line in Figure~\ref{fig:periodograms}), for which we measure an amplitude of
$0.36\,\mathrm{m\,s}^{-1}$, lies in a region where the completeness is only $\sim$45\%, which is consistent with recovering
tentative, but inconclusive statistical evidence for such a signal (see red crosses in Fig.s~\ref{fig:conflevel} and
\ref{fig:detlim}).

The same experiment performed over the GJ~699 time baseline resulted in a more conservative detection limit
($\sim1.20\,\mathrm{m\,s}^{-1}$). That crucial impact on the detectability of the signal is the result of having three nights
of high-cadence data instead of four, and of having consecutive or non-consecutive observing nights that tangle the window
function. In other words, even when both GJ~588 and GJ~699 are comparable in terms of stability, the more complicated window function of
GJ~699 as a result of the sampling cadence might have prevented us from reaching higher sensitivity.

\section{Discussion and Conclusions} \label{sec:discussion} 
Theoretical studies predict that main sequence M dwarfs can oscillate with periods ranging from 20~min up to 3~h. The detection
of such pulsations will open the whole field of asteroseismology for this spectral type; but first stellar pulsation on M
dwarfs have to be observationally confirmed. This is one of the goals of the Cool Tiny Beat program (CTB), which makes use
of the high-precision spectroscopy given by HARPS and HARPS-N to explore the short-time domain of a sample of M dwarfs with
high-cadence observations. 

The CTB thorough monitoring of the night using high-cadence observations deepens into a time domain which has not yet been
widely explored by HARPS. As a result, we had to deal with some unexpected and not very well known instrumental effects. In this study, we
detailed the main corrections to be applied to deal with such effects and thus we set up a procedure for the analysis of the CTB data or in general, for the analysis of high-cadence, high-precision Doppler time-series. We presented here the first results of the survey in search for stellar pulsations in M dwarfs. In particular, we
focused on GJ~588 and GJ~699 (Barnard's star), two of the most long-term stable stars targets of the sample (i.e. with no
planet or strong activity reported so far) for which CTB collected four and three whole data nights, respectively. 

Even when no signals compatible with pulsations were detected above the classical confidence thresholds (FAP=1\% in the case of the likelihood periodograms, or 4~$\sigma$ in the case of the power spectra), we detected some excess of power in the periodograms and their power spectra for the two targets. More and higher precision data would be needed to confirm or refute them.

Giving serious thought to the fact that the signals could be caused by stellar oscillations, we have checked their compatibility with pulsation models. Results indicate that the putative signal at 12~d$^{-1}$ ($\sim$2~h) found for GJ~588
would be compatible with low-radial, low degree $\ell$=1 and $\ell$=2 \emph{g}-modes produced in stellar theoretical models also compatible with GJ~588 physical parameters in terms of mass, age, $T_\mathrm{eff}$ and $\log{g}$. On the contrary,
GJ~699 was found not to be simultaneously compatible with any excited model in terms of its physical parameters and the periods of its putative signals.

Finally, we derived an amplitude detection limit for the detection of pulsations in M dwarfs with HARPS. Results indicated that no
signal below $\sim0.5\mathrm{m\,s}^{-1}$ can be detected with a confidence level better than 90\% on the most Doppler stable
M-dwarf studied so far (GJ~588). To obtain this limit, we used the standard CTB observational strategy for the pulsations
science case. The higher threshold derived for GJ~699 --for which we only get three non-consecutive nights-- demonstrates the
crucial impact of the observational cadence in the detectability of a pulsation signature. The success of any spectroscopic
program searching for pulsations will ultimately rely on even higher precision, but also on optimal sampling strategies.

Before solar-like oscillations were finally detected, many studies on different spectral type stars reported hints of power
excess. This was the case of the G2V star $\alpha$~Cen~A, for which several studies \citep[e.g.][]{schou2000} reported an excess of power before solar-like oscillations were finally confirmed by \cite{bouchy2001, bouchy2002} using the spectrometer CORALIE. This final confirmation came hand in hand with the rapid improvement of spectrographs, which aimed at detecting the first exoplanets. This could also end up being the case for M dwarfs pulsations. If the theoretical studies are accurate and the driving mechanisms can efficiently develop oscillations in M dwarf stars either: i) the amplitudes are very low thus its confirmation requires of more precise
spectrographs (e.g. the forthcoming ESPRESSO/VLT; \citealt{pepe2010}, or HIRES/E-ELT; \citealt{zerbi2014}) or, ii) the size of the sample combined with a possible non-pure instability region is preventing us from having an observational
confirmation (e.g. only $\sim$~40\% of the $\delta$ Scuti within its instability region present oscillations \citealt{balona2011}). 

In spite of this result, the ``Cool Tiny Beats'' survey takes us one step closer to the observational detection of M dwarfs
pulsations and illustrates the challenges of high precision experiments, even with current state-of-the-art instrument like
HARPS. Besides enlarging our observed sample with HARPS and HARPS-N, we plan to extent the search to ESPRESSO in the near
future to monitor the two islands of instability predicted by RL14. Our aim is to build up a more targeted asteroseismology sample where models
predict pulsations to be more conspicuous. 

\section*{acknowledgements}
We acknowledge funding from AYA2014-30147-C03-01 by MINECO/Spain. This study is based on observations made with the 3.6~m ESO Telescope at la Silla under programme ID 191.C-0505. We acknowledge Stefan Dreizler,  Enrico Gerlach, Sandra Jeffers, James Jenkins, Christofer Marvin, Julien Morin, Aviv Ofir, Ansgar Reiners and Ulf Seemann their participation in the preparation of this programme proposal and as long as their support and useful discussions. The authors thank the referee T. B\"ohm for his suggestions that helped improved this paper.
\bibliography{masterbibzaira}
\bibliographystyle{mn2e}
\label{lastpage}
\begin{appendix}
\section{Detailed Frequentist analysis} \label{app:A}

	\begin{figure*}
	\includegraphics[width=0.84\textwidth]{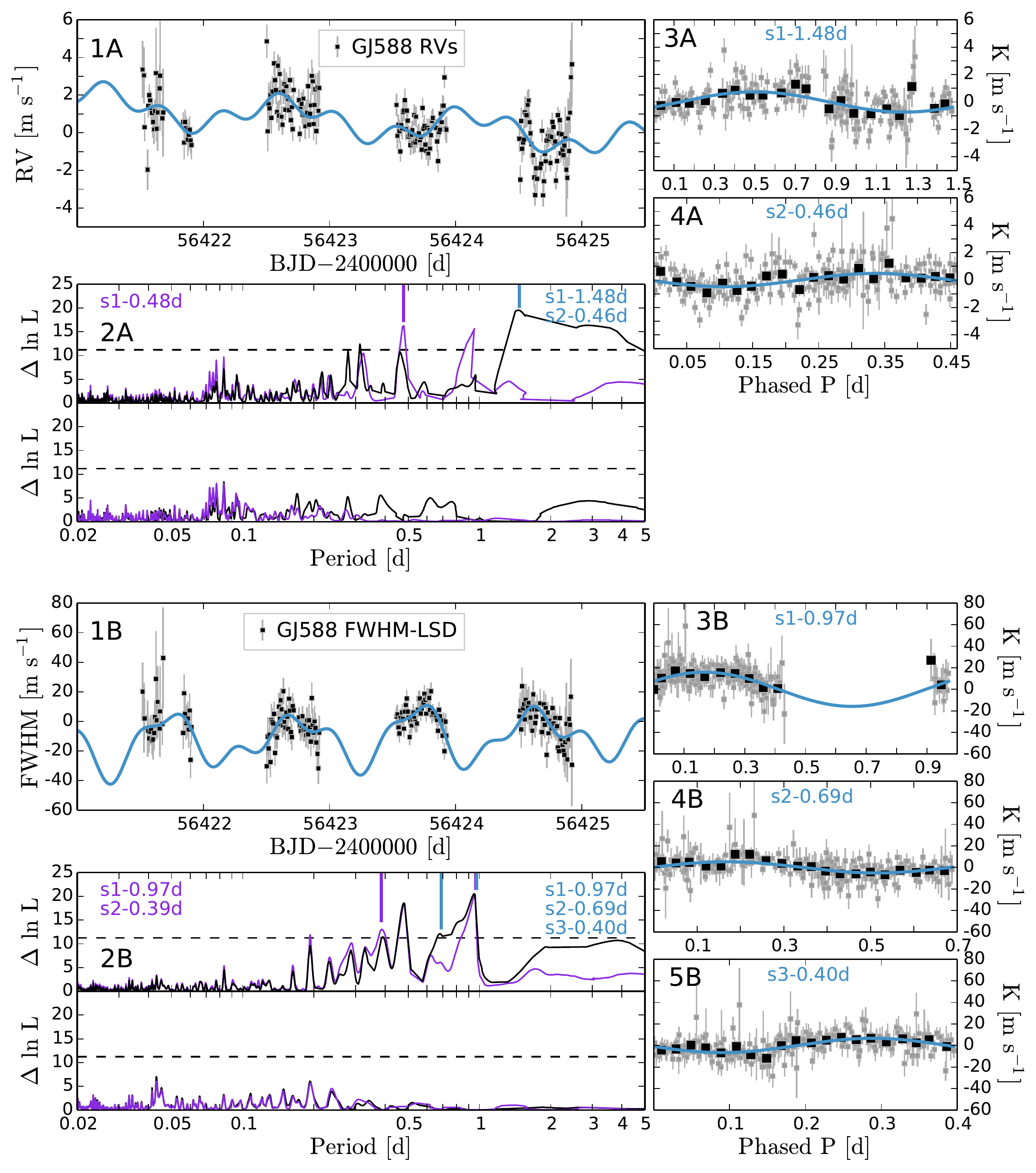}
	\caption{Periodic solutions for GJ~588 RVs (A panels) and FWHM-LSD (B panels) indices detected beyond the pulsational range ( P~$>3$~h, or P~$>0.125$~d). Instrumental distortions characteristic of the night (e.g. chromatic seeing or atmospheric dispersion among others) cause 1-day signals and sub-multiples. The light blue line plotted in panels 1A and 1B corresponds to the best model fitted to the observations (black squares) when we consider all the data as an single dataset (see in panels 2A and 2B the corresponding periodograms in black). For the RVs, such model includes two sinusoid (P~$_{1}\sim$1.5~d, P~$_{2}\sim$0.5~d), while for the FWHM-LSDs the model includes three sinusoids (P~$_{1}\sim$1~d, P~$_{2}\sim$0.7~d and P~$_{2}\sim$0.4~d). Panels 3A, 4A and 3B, 4B and 5B show the data (grey squares) phased folded to each of these periodic signals (light blue lines). For the sake of clarity, bins of the observations are show with black squares. The periods of these signals are highlighted with light blue vertical lines in panels 2A and 2B. Purple periodograms in 2A and 2B panels result when we consider each night as an independent dataset. This approach  works as a filter for periods lower than 1-day. Regardless of the method used, we recover the same peaks in the range of study (20~min to 3~h), but we need simpler models with less sinusoids when we treat the nights independently (purple periodograms).}
	\label{fig:allgj588}
	\end{figure*}

	\begin{figure*}
	\includegraphics[width=0.84\textwidth]{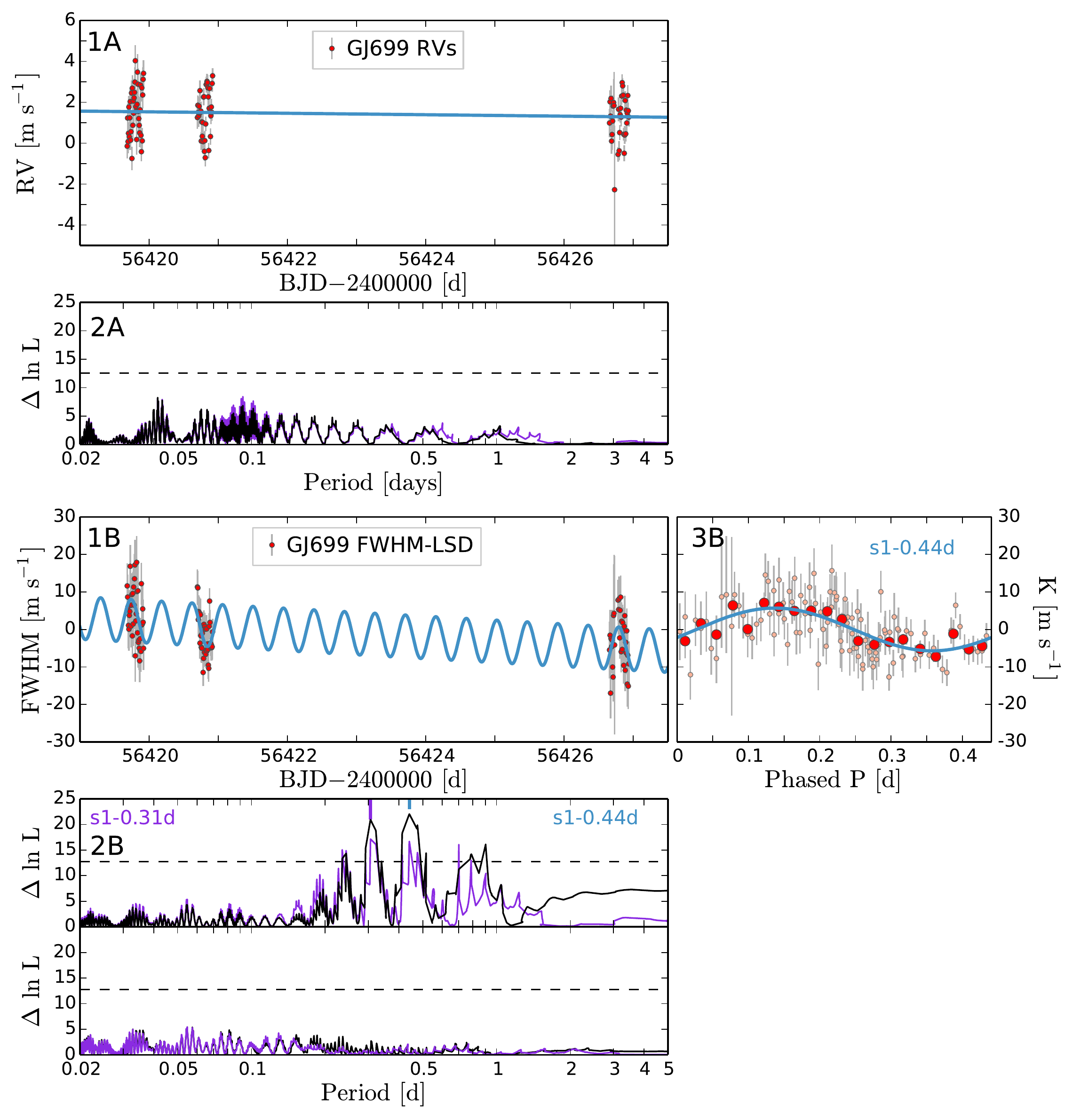}
	\caption{Periodic solutions for GJ~699 RVs (A panels) and FWHM-LSD (B panels) found in the P~$>3$~h (or P~$>0.125$~d) range. Panels 1A and 1B show the  time-series of the observations (red dots) and the best fitted model (blue lines). Panels 2A and 2B show the likelihood periodograms. The black periodograms correspond to the analysis of the data as a single dataset. The purple periodograms correspond to the analysis of the observations took at different nights as independent datasets. The blue lines in all panels correspond to models obtained in the case of a single dataset analysis. Only one sinusoid had to be included for the FWHM-LSDs (P~$\sim0.4$~d in case of a single dataset analysis, and P~$\sim0.3$~d in case of treating the nights independently). The panel 3B shows the observations phased folded to P~$\sim0.4$~d period. The big red dots are bins of the observations, which are indicated with smaller symbols.}
	\label{fig:allgj699}
	\end{figure*}
	
\end{appendix}
\end{document}